\documentclass[epj]{svjour}
\usepackage{graphics,enumerate}
\usepackage{cite, hyperref}
\usepackage{ulem, color}
\usepackage{amsmath,amssymb}
\usepackage[utf8]{inputenc}
\usepackage{color}

\def\CF{\mathrm{C_F}}
\def\CFsq{\mathrm{C_F^2}}
\def\CA{\mathrm{C_A}}
\def\CAsq{\mathrm{C_A^2}}
\def\Nc{\mathrm{N_c}}
\renewcommand{\d}{\mathrm{d}}

\def\A{\mathcal{A}}

\def\B{\mathcal{B}}

\def\cC{\mathcal{C}}
\def\inn{\mathrm{in}}

\def\NG{\mathrm{NG}}
\def\as{\alpha_s}
\def\asb{\bar{\alpha}_s}
\def\cF{\mathcal{F}}
\def\cG{\mathcal{G}}
\def\cO{\mathcal{O}}
\def\cW{\mathcal{W}}
\def\cWb{\overline{\mathcal{W}}}
\def\cS{\mathcal{S}}
\def\cR{\mathcal{R}}
\def\R{{\scriptscriptstyle\mathrm{R}}}
\def\V{{\scriptscriptstyle\mathrm{V}}}

\def\bT{\mathbf{T}}

\begin{document}

\title{Jet mass distribution in Higgs/vector boson + jet events at hadron colliders with ${\boldsymbol k}_{\boldsymbol t}$ clustering}
\author{Na\"{i}ma Ziani\inst{1}\fnmsep\thanks{\email{naima.ziani@univ-batna.dz}}\and Kamel Khelifa-Kerfa\inst{2}\fnmsep\thanks{\email{kamel.kkhelifa@iu.edu.sa}}\and
Yazid Delenda\inst{1}\fnmsep\thanks{\email{yazid.delenda@univ-batna.dz} (corresponding author)}}
\institute{Laboratoire de Physique des Rayonnements et de leurs Int\'{e}ractions avec la Mati\`{e}re,\\
D\'{e}partement de Physique, Facult\'{e} des Sciences de la Mati\`{e}re,\\
Universit\'{e} de Batna--1, Batna, Algeria\and
Department of Physics, Faculty of Science, Islamic University of Madinah,\\
Madinah 42351, Saudi Arabia }

\date{
}

\abstract{We address the issues of clustering and non-global logarithms for jet shapes in the process of production of a Higgs/vector boson associated with a single hard jet at hadron colliders. We perform an analytical fixed-order calculation up to second order in the coupling as well as an all-orders estimation for the specific invariant mass distribution of the highest-$p_t$ jet, for various jet algorithms. Our results are derived in the eikonal (soft) limit and are valid up to next-to-leading logarithmic accuracy. We perform a matching of the resummed distribution to next-to-leading order results from \texttt{MCFM} and compare our findings with the outputs of the Monte Carlo event generators \texttt{Pythia 8} and \texttt{Herwig 7}. After accounting for non-perturbative effects we compare our results with available experimental data from the CMS collaboration for the $Z$ + jet production. We find good agreement over a wide range of the observable.
}

\authorrunning{N. Ziani et al.}
\titlerunning{Jet mass distribution in $H/V$ boson + jet events at hadron colliders with $k_t$ clustering}

\maketitle

\section{Introduction}
\label{intro}

The invariant mass of a jet is a typical example of a jet shape that plays an important role in the study of the substructure of jets, testing QCD, and identifying new-physics signals. Being sensitive to soft and/or collinear emissions from the parton initiating the jet and from the other incoming and outgoing partons, this observable provides an indispensable mean for probing various aspects that are relevant to achieving better accuracy in QCD calculations. Examples of such aspects include, on the non-perturbative side, hadronisation corrections, underlying event, pile-up interactions, and on the perturbative side, initial and final-state radiation, colour flow, resummation of large logarithms, etc. Analytical calculations for these aspects pave the way for a deeper insight into QCD processes, a better control of theoretical uncertainties, and a precise quantification of missing higher-order contributions and their significance, all of which are issues not very clear in Monte Carlo event generators.

In this paper we shed light on the resummation of large logarithms that arise due to a miscancellation of soft and collinear singularities between real emissions and their corresponding virtual corrections. The convergence of the perturbative series, in the invariant jet mass ($m_j$) distribution, is spoiled by the presence of large logarithms in the ratio of the jet mass and its transverse momentum $p_t$, $L=\ln(m_j/p_t)$, at each order in perturbation theory. In the exponent of the integrated distribution, these logarithms take the form $\alpha_s^nL^m$, with $\alpha_s$ being the strong coupling constant and $m\leq(n+1)$, and thus they require an all-orders resummation. A next-to-leading logarithmic (NLL) resummation ensures that all single logarithms of the form $\alpha_s^nL^n$ are resummed, in addition to the leading (double) logarithms (LL) $\alpha_s^n L^{n+1}$.

The jet mass is a non-global observable, i.e., an exclusive observable that is sensitive only to gluon emissions which end up inside the jet. To ensure a proper NLL resummation then its distribution must carefully be treated for a class of large single logarithms known as non-global logarithms (NGLs), which are related to secondary non-Abelian emissions of soft gluons \cite{Dasgupta:2001sh,Dasgupta:2002bw}. Furthermore, another type of large single logarithms known as clustering logarithms (CLs) \cite{Banfi:2005gj,Delenda:2006nf}, related to primary gluon emissions off the hard Born configuration, needs to be resummed when the jets are reconstructed using jet algorithms such as $k_t$ \cite{Catani:1993hr,Ellis:1993tq} and Cambridge-Aachen (C-A) \cite{Dokshitzer:1997in,Wobisch:1998wt}. The anti-$k_t$ clustering algorithm \cite{Cacciari:2008gp} is known to cause no CLs (see for instance refs. \cite{Banfi:2010pa,Dasgupta:2012hg}). The full resummation of both NGLs and CLs has thus far proven to be a formidable challenge. The resummation of NGLs is usually achieved numerically via a Monte Carlo approach \cite{Dasgupta:2001sh, Dasgupta:2002bw} in the large-$\Nc$ limit ($\Nc$ being the number of quark colours), though full-colour numerical resummation has been provided in refs. \cite{Hatta:2013iba, Hagiwara:2015bia} based on an analogy between small-$x$ BFKL resummation in Regge scattering and the Weigert equation \cite{Weigert:2003mm}. Additionally, NGLs may also be resummed via an evolution equation known as the Banfi-Marchesini-Smye (BMS) equation \cite{Banfi:2002hw} valid at large $\Nc$.

In this work we study the jet mass distribution in the process of production of a single jet associated with a vector boson ($\gamma$, $Z$ or $W$) or a Higgs boson $H$ at the Large Hadron Collider (LHC). In ref. \cite{Dasgupta:2012hg}, the jet mass distribution was calculated at NLL accuracy combined with next-to-leading order (NLO) results in $Z$ + jet and di-jet processes at hadron colliders, \footnote{Note that our convention for LO, NLO, etc, is different from that in ref. \cite{Dasgupta:2012hg}. In our convention, the LO differential distribution is proportional a delta function.} for jets defined with the anti-$k_t$ clustering algorithm. The NGLs were computed therein analytically at fixed order (at $\cO(\as^2)$) and numerically to all orders in the large-$\Nc$ approximation. In the context of soft-collinear effective theory the jet mass distribution was also studied in ref. \cite{Liu:2014oog} for di-jet events, in ref. \cite{Chien:2012ur} for $\gamma$ +jet events, and in ref. \cite{Jouttenus:2013hs} for $H$ + jet events. We elaborate herein on the work of ref. \cite{Dasgupta:2012hg} by considering the jet mass distribution when jets are reconstructed using $k_t$ or C-A clustering algorithms. We additionally consider other vector bosons, namely $\gamma$ and $W$, as well as Higgs boson + jet production processes. On the experimental side, the jet mass distribution in $W/Z$ + jet events at the LHC was studied by the CMS collaboration \cite{Chatrchyan:2013rla}, where the jets were reconstructed using various jet algorithms. Additional jet substructure techniques such as trimming, filtering, and pruning, were also addressed in the same work \cite{Chatrchyan:2013rla}. We do not address these techniques in the present paper.

We compute NGLs and CLs at fixed order, specifically at $\mathcal{O}(\alpha_s^2)$ where they first appear, for the invariant mass distribution of the highest-$p_t$ jet. We provide results for the following three jet algorithms: $k_t$, C-A and anti-$k_t$, where we note that for the latter algorithm NGLs were first computed in ref. \cite{Dasgupta:2012hg} and that CLs are absent. Moreover, we approximate the all-orders resummed CLs and NGLs by an exponential of the $\mathcal{O}(\alpha_s^2)$ result in the case of $k_t$ and C-A algorithms. This is justified by the fact that for the anti-$k_t$ algorithm the said exponential approximates the all-orders numerical result very well as we shall demonstrate. \footnote{While the all-orders numerical resummation of NGLs for the anti-$k_t$ algorithm may be computed using the Monte Carlo code of ref. \cite{Dasgupta:2001sh}, as was done in ref. \cite{Dasgupta:2012hg}, we found that this code produces unreliable results for some dipoles in the case of $k_t$ clustering. Note that the C-A algorithm is not implemented in the code of ref. \cite{Dasgupta:2001sh}.} We then compare the NLL-resummed and NLO-matched result for the jet mass distribution, which includes the resummed global and non-global (NGLs and CLs) form factors convoluted with the Born cross-section and corrected for NLO effects for each of the four $V/H$ + jet processes, with results from \texttt{Pythia 8} \cite{Sjostrand:2014zea} and \texttt{Herwig 7} \cite{Bahr:2008pv,Bellm:2015jjp} parton showers. Finally, we estimate the non-perturbative corrections to this distribution and compare our predictions with experimental data from the CMS collaboration \cite{Chatrchyan:2013rla} for the jet mass distribution in $Z$ + jet events at the LHC.

This paper is organised as follows. In section 2 we discuss kinematics of the processes under consideration and define our observable. We calculate, in section 3, the distribution of the jet mass at leading order and construct the resummed global form factor up to NLL accuracy in the exponent. In section 4 we compute the leading CLs at $\mathcal{O}(\alpha_s^2)$ for both $k_t$ and C-A clustering algorithms, which happen to give identical results at this particular order. We also calculate NGLs at $\mathcal{O}(\alpha_s^2)$ for the aforementioned jet algorithms in addition to the anti-$k_t$. We are then able to assess the impact of the various clustering algorithms on NGLs. In section 5 we discuss the all-orders resummation of NGLs and CLs. In section 6 we compare our NLL-resummed result including NLO corrections for the jet mass distribution with the outputs of \texttt{Pythia 8} and \texttt{Herwig 7} parton showers. In section 7 we estimate the non-perturbative corrections, which include hadronisation corrections and the underlying event, on the distribution, and compare our results with the experimental data. Finally, in section 8, we draw our conclusions.

\section{Setup}

\subsection{Processes and kinematics}

In this paper we are interested in the calculation of both CLs and NGLs at single logarithmic accuracy, for the jet mass distribution in the process of production of a single jet associated with a vector ($W/Z/\gamma$) or Higgs boson at hadron colliders. For this purpose, it suffices to consider the eikonal (soft) approximation in the squared matrix elements for the emission of gluons. The emitted gluons are assumed to be strongly ordered in transverse momenta, i.e., $k_{tn}\ll\cdots\ll k_{t2}\ll k_{t1}\ll p_t$, where $k_{ti}$ is the transverse momentum of the $i^{\mathrm{th}}$ emission and $p_t$ is that of the outgoing hard jet. The latter ordering simplifies the calculations of the emission amplitudes while being sufficient for capturing the single logarithmic CLs and NGLs.

For a vector boson + one jet production in hadron collisions, there are three partonic channels that contribute to the Born process, namely: $q\bar{q}\to g$, $qg\to q$, and $\bar{q}g\to\bar{q}$. For $W^{\pm}$ production, flavour changing needs to be taken into account at the Born level, but this does not affect the QCD structure of initial and final-state radiation. As for the Higgs + one jet process there are four partonic channels to be considered. These are: $qg\to qH$, $\bar{q}g\to\bar{q}H$, $q\bar{q}\to Hg$, and $gg\to Hg$. As far as QCD calculations are concerned all mentioned channels, whether for Higgs or vector boson production, are in fact identical as they all involve three hard coloured (QCD) partons and a colour-neutral boson. This means that the resummation of the jet mass distribution is essentially identical in all of the said channels, with differences pertaining to just the Born cross-section and the associated colour factors for the various channels. We note that the relevant total cross-sections have been calculated up to next-to-next-to-leading order (NNLO): Higgs + jet in refs. \cite{Boughezal:2015dra, Boughezal:2015aha, Caola:2015wna, Chen:2016zka}, $Z$ + jet in refs. \cite{Ridder:2015dxa, Boughezal:2015ded}, $W$ + jet in ref. \cite{Boughezal:2015dva}, and $\gamma$ + jet in ref. \cite{Campbell:2016lzl}.

\begin{figure}[ht]
\centering
\resizebox{0.45\textwidth}{!}{\includegraphics{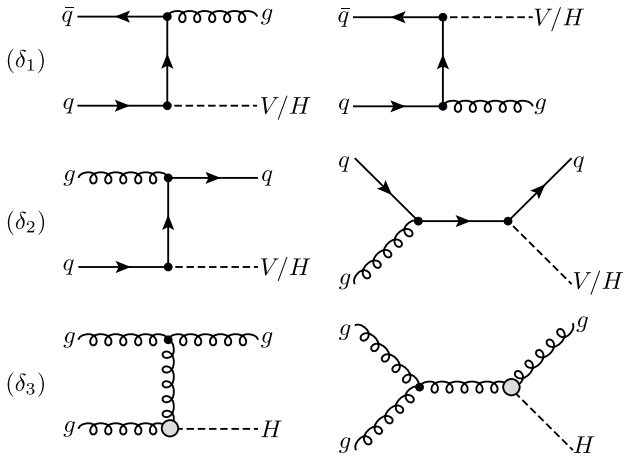}}
\caption{\label{fig:feyn}Partonic processes considered in this paper.}
\end{figure}

In the current work, we henceforth consider the three partonic channels shown in figure \ref{fig:feyn}: $(\delta_1):q\bar{q}\to g+X$, $(\delta_2):qg\to q+X$, and $(\delta_3):gg\to g+X$, \footnote{For our QCD calculations, the channel $\bar{q}g\to\bar{q}$ is in fact identical to $qg\to q$.} where $X$ refers to the colour-neutral boson ($\gamma$, $Z$, $W^{\pm}$ or $H$).
We label the incoming partons with $(a)$ and $(b)$ and the outgoing parton initiating the hard jet with $(j)$. The four-momenta of the three hard Born partons and the emitted soft gluons are given by
\begin{subequations}
\begin{align}\label{eq:FourMomenta_Definition}
p_a&=x_a\,\frac{\sqrt{s}}{2}\left(1,0,0,1\right),\\
p_b&=x_b\,\frac{\sqrt{s}}{2}\left(1,0,0,-1\right),\\
p_j&=p_t\left(\cosh y,\cos\varphi,\sin\varphi,\sinh y\right),\\
k_i&=k_{ti}\left(\cosh\eta_i,\cos\phi_i,\sin\phi_i,\sinh\eta_i\right),
\end{align}
\end{subequations}
where $\eta_i$ and $\phi_i$ are the rapidity and azimuth of the $i^\mathrm{th}$ emission and $y$ and $\varphi$ are those of the outgoing hard jet, measured with respect to the beam axis. The incoming partons $a$ and $b$ carry momentum fractions $x_a$ and $x_b$ of the incoming protons, and $\sqrt{s}$ is the collision centre-of-mass energy. We shall be ignoring recoil against soft emissions throughout, as it is beyond single logarithmic accuracy.

\subsection{Jet mass observable and jet algorithms}

We study the normalised (squared) invariant mass of the outgoing hard jet $j$ defined by
\begin{equation}\label{eq:def}
\begin{split}
\varrho&\equiv\frac{1}{p_t^2}\left(p_j+\sum_{i\in\mathrm{jet}}k_i\right)^2\approx\frac{2}{p_t^2}\sum_{i\in\mathrm{jet}}k_i\cdot p_j=\sum_{i\in\mathrm{jet}}\varrho_i\,,\\
\varrho_i&\equiv\,\frac{2\,k_i\cdot p_j}{p_t^2}=2\,\frac{k_{ti}}{p_t}\left[\cosh(\eta_i-y)-\cos(\phi_i-\varphi)\right],
\end{split}
\end{equation}
where the sum is over all emitted soft gluons which end up inside the hard jet after the application of a jet algorithm on the final state partons. Notice that we are considering massless quarks and that the soft approximation has been assumed in the above equation whereby $p_j\cdot k_i\gg k_\ell\cdot k_m$.

The $k_t$, C-A and anti-$k_t$ jet algorithms work as follows. For each pair $(im)$ of hadrons in the final state one defines a distance
\begin{subequations}\label{eq:AlgDistances}
\begin{align}
d_{im}=\min(k_{ti}^{2p},k_{tm}^{2p})\left[(\eta_i-\eta_m)^2+(\phi_i-\phi_m)^2\right],
\end{align}
and for each single hadron a beam distance
\begin{align}
d_i=k_{ti}^{2p}\,R^2\,,
\end{align}
\end{subequations}
for some fixed jet radius parameter $R$. Here the parameter $p=1,0,-1$ for $k_t$, C-A, and anti-$k_t$ clustering, respectively. If the smallest of all of these distances is $d_{im}$, then particles $i$ and $m$ are combined into a single particle with four-momentum $p_i+p_m$, whereas if the smallest is $d_i$ then particle $i$ is considered as a jet and is removed from the list of particles. This procedure is iterated until one is left only with jets in the final state.

For the $k_t$ algorithm, and in the regime of strongly-ordered emissions, the clustering of particles starts with the softest real gluon. Then, in a given event this softest gluon is dragged towards the next-to-softest real parton within a circle of radius $R$ in the $(\eta,\phi)$ plane. If no such harder parton exists then this softest gluon is considered as a jet and is removed from the list of partons. The process is then repeated until no particles are left. When clustering two partons together, the resulting pseudo-jet is essentially aligned along the direction of the harder, and its four-momentum is just that of the harder parton.

For the anti-$k_t$ algorithm, on the other hand, clustering starts with the hardest particle, and hence it works in an apposite way to $k_t$ clustering. For the C-A algorithm, only geometric distances between partons in the $(\eta,\phi)$ plane decide how clustering happens. Particles which are closest to each other get clustered first.

\subsection{Jet mass distribution}

In what follows we calculate at NLL accuracy the jet mass distribution for a given channel $\delta$, defined by (following the notation of refs. \cite{Dasgupta:2012hg, Banfi:2004yd})
\begin{equation}\label{eq:dff}
\frac{\d\Sigma_\delta(\rho)}{\d\B_\delta}=\int_0^\rho\frac{\d^2\sigma_\delta}{\d\B_\delta\,\d\varrho}\,\d\varrho\,,
\end{equation}
where $\d^2\sigma_\delta/\d\B_\delta\,\d\varrho$ is the differential cross-section with respect to both the Born configuration $\B_\delta$ and the jet mass observable $\varrho$. Details of the differential Born configuration $\d\B_\delta$ are discussed further in appendix \ref{sec:Born}. The integrated jet mass distribution is obtained by integrating $\d\Sigma_\delta(\rho)/\d\B_\delta$ over $\B_\delta$ with some chosen kinematical cuts (which we denote by $\Xi_\B$), and summing over all Born channels. That is
\begin{align}\label{eq:Integrated_Sogma_delta}
\Sigma(\rho)=\sum_{\delta}\int\d\B_\delta\,\frac{\d\Sigma_\delta(\rho)}{\d\B_\delta}\,\Xi_\B\,.
\end{align}

Following ref. \cite{Dasgupta:2012hg}, we write eq. \eqref{eq:dff} in the region $\rho\ll1$ in the factorised form
\begin{align}\label{eq:dSigma}
\frac{\d\Sigma_\delta(\rho)}{\d\B_\delta}=\frac{\d\sigma_{0,\delta}}{\d\B_\delta}\,f_{\B,\delta}(\rho)\,C_{\B,\delta}(\rho)\,,
\end{align}
where $\d\sigma_{0,\delta}/\d\B_\delta$ is the differential partonic Born cross-section for channel $\delta$ (see appendix \ref{sec:Born}) and the factor $C_{\B,\delta}(\rho)$ depends on the Born kinematics and has the perturbative expansion
\begin{equation}
C_{\B,\delta}(\rho)=1+\as\,C_{\B,\delta}^{(1)}(\rho)+\as^2\,C_{\B,\delta}^{(2)}(\rho)+\cdots\,,
\end{equation}
where $C_{\B,\delta}^{(n)}(\rho)$ are channel-dependent terms that correct the resummation for non-logarithmically-enhanced terms. The $\rho$-dependent function $f_{\B,\delta}(\rho)$ resums all the large logarithms. It has the form \cite{Banfi:2004yd}
\begin{equation}\label{eq:fB}
f_{\B,\delta}(\rho)=\exp\left[L\,\tilde{g}_1(\as L)+\tilde{g}_2(\as L)+\as\,\tilde{g}_3(\as L)+\cdots\right],
\end{equation}
where the function $L\,\tilde{g}_1$ resums the leading (double) logarithms (LL) of the form $\as^nL^{n+1}$, $\tilde{g}_2$ resums next-to-leading (single) logarithms (NLL) of the form $\as^nL^n$, and $\as\,\tilde{g}_3$ resums next-to-next-to-leading logarithms (NNLL) of the form $\as^nL^{n-1}$, and so on, with $L=\ln(R^2/\rho)$. The LL function $\tilde{g}_1$ receives contributions from soft-collinear emissions from the parton initiating the jet and depends on its colour Casimir scalar. The NLL function $\tilde{g}_2$ receives contributions from various sources:
\begin{enumerate}[(a)]
\item hard-collinear emissions from the outgoing hard parton,
\item soft wide-angle emissions from all hard partons,
\item starting at $\mathcal{O}(\as^2)$, NGLs from soft wide-angle correlated secondary emissions, and
\item CLs, when jet algorithms other than anti-$k_t$ are implemented for jet reconstruction, from soft wide-angle primary emissions off the hard partons. These again appear starting from $\mathcal{O}(\as^2)$.
\end{enumerate}

The whole function $\tilde{g}_1$ and parts of $\tilde{g}_2$, namely contributions (a) and (b) stated above, have been determined in ref. \cite{Dasgupta:2012hg} for the anti-$k_t$ algorithm. The exact same result also applies for the case of $k_t$ and C-A clustering as the effect of jet algorithms first appears at $\mathcal{O}(\as^2)$. Our task is to determine the other two contributions to $\tilde{g}_2$, namely (c) NGLs and (d) CLs, for $k_t$ and C-A algorithms. Before doing so, we review in the next section the basic calculations that lead to the determination of $\tilde{g}_1$ and contributions (a) and (b) of $\tilde{g}_2$.

\section{One-gluon emission}

\subsection{Fixed-order calculation}
\label{sec:fo}

In this section we compute the jet mass distribution at leading order in QCD and present the all-orders resummed result. Our calculations are valid in the eikonal approximation and accurate up to NLL accuracy. First, we define the following antenna functions relevant for the squared matrix elements for the emission of soft gluons
\begin{subequations}
\begin{align}\label{eq:Antennas_defs}
w_{\alpha\beta}^i&=\frac{k_{ti}^2}{2}\,\frac{p_\alpha\cdot p_\beta}{(p_\alpha\cdot k_i)\,(k_i\cdot p_\beta)}\,,\\
\A_{\alpha\beta}^{ik}&=w^i_{\alpha\beta}\left(w^k_{\alpha i}+w^k_{i\beta}-w^k_{\alpha\beta}\right).
\end{align}
\end{subequations}
It is worth noting that these antennae are purely angular functions, i.e., they involve no energy or momentum dependence.

\begin{figure}[ht]
\centering
\resizebox{0.45\textwidth}{!}{\includegraphics{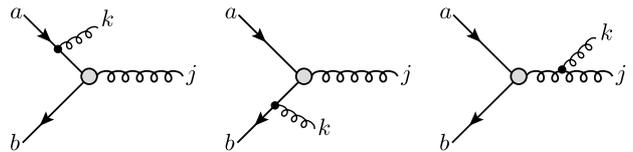}}
\caption{\label{fig:1loop}Feynman diagrams for the emission of one gluon off the three-hard-partons configuration $(abj)$.}
\end{figure}

Consider the process of emission of a single soft gluon off the three-hard-legs Born configuration $(abj)$, i.e., the process $a+b\to j+k_1$ shown in figure \ref{fig:1loop}.
The corresponding factorised eikonal amplitude squared is given by
\begin{equation}\label{eq:W1R}
\cW_{1,\delta}^\R=\sum_{(i\ell)\in\Delta_\delta}\mathcal{C}_{i\ell}\,w_{i\ell}^1\,,
\end{equation}
with $\Delta_\delta=\{(ab),(aj),(bj)\}$ denoting the three dipoles formed from the partons in channel $\delta$. The colour factor $\mathcal{C}_{i\ell}$ is defined as
\begin{equation}\label{eq:ColorFactor}
\cC_{i\ell}=-2\,\bT_i\cdot\bT_\ell\,,
\end{equation}
where $\bT_i$ are the generators of the $\mathrm{SU}(\Nc)$ group with Casimir scalar given by $\bT_i^2=\CF=(\mathrm{N_c^2}-1)/(2\,\Nc)$ for quarks (and anti-quarks) and $\bT_i^2=\CA=\Nc$ for gluons. Conservation of colour implies that for our leading-order process $a+b\to j+k_1$ we have \cite{Delenda:2015tbo}: $\bT_a+\bT_b+\bT_j=0$, where the generators are taken as if all partons were incoming. Explicitly written, the colour factors relevant to our dipoles are: $\mathcal{C}_{q\tilde{q}}=2\,\CF-\CA=-1/\Nc$ and $\mathcal{C}_{qg}=\mathcal{C}_{gg}=\CA=\Nc$.

The term $\cW_{1,\delta}^\R$ is the eikonal amplitude squared for the emission of a {\it real} soft gluon in the partonic sub-process $\delta$. The corresponding virtual correction in the eikonal limit is simply $\cW_{1,\delta}^\V = -\cW_{1,\delta}^\R$. Notice that we are adopting the notation used in our previous work on eikonal amplitudes for $e^+e^-\to$ di-jet process \cite{Delenda:2015tbo}. In our recent paper \cite{Khelifa-Kerfa:2020nlc} we have generalised the latter to the case of hadron collisions, specifically considering three-hard-legs Born processes. The corresponding phase-space factor is given by
\begin{equation}\label{eq:dPhi1}
\d\Phi_1=\asb\,\frac{\d\xi_1}{\xi_1}\,\d\eta_1\,\frac{\d\phi_1}{2\pi}\,,
\end{equation}
where $\bar{\alpha}_s=\alpha_s/\pi=g_s^2/4\pi^2$, $g_s$ is the strong coupling, and $\xi_1=k_{t1}/p_t$. The running of the coupling is irrelevant at one loop and only becomes important at higher orders. The full resummation that we present later will include running-coupling effects.

Following the procedure of measurement operators (see for instance ref. \cite{Khelifa-Kerfa:2015mma}), we write the jet mass distribution at one loop as
\begin{equation}\label{eq:dSigma_1loop}
f_{\B,\delta}^{(1)}(\rho)=-\int\d\Phi_1\,\cW_{1,\delta}^\R\,\Theta(\varrho_1-\rho)\,\Xi_{\inn}(k_1)\,,
\end{equation}
where the function $\Xi_{\inn}(k_1)$ ensures that the angular integration region for gluon $k_1$ is such that it gets clustered to the hard jet when the jet algorithm is applied. At this order all jet algorithms essentially work in the same manner, and $\Xi_{\inn}(k_1)$ is then a simple Heaviside step function; $\Xi_{\inn}(k_1)=\Theta_{\inn}(k_1)=\Theta\left[R^2-(\eta_1-y)^2-(\phi_1-\varphi)^2\right]$. At higher loops, as we shall see, this is not as simple.

Substituting the expression of the eikonal amplitude squared \eqref{eq:W1R} into eq. \eqref{eq:dSigma_1loop} we obtain
\begin{align}
f_{\B,\delta}^{(1)}(\rho)&=-\sum_{(i\ell)\in\Delta_\delta}\mathcal{C}_{i\ell}\,\asb\int\frac{\d\xi_1}{\xi_1}\,\d\eta_1\,\frac{\d\phi_1}{2\pi}\,w_{i\ell}^1\,\times\notag\\
&\times\Theta(\varrho_1-\rho)\,\Theta[R^2-(\eta_1-y)^2-(\phi_1-\varphi)^2]\,,\label{eq:twn}
\end{align}
with the antenna function $w_{i\ell}^1$ for each emitting dipole in $\Delta_\delta$ given by
\begin{subequations}
\begin{align}
w_{ab}^1&=1\,,\\
w_{aj}^{1}&=\frac{1}{2}\,\frac{\exp(\eta_1-y)}{\cosh(\eta_1-y)-\cos(\phi_1-\varphi)}\,,\\
w_{bj}^{1}&=\frac{1}{2}\,\frac{\exp(y-\eta_1)}{\cosh(\eta_1-y)-\cos(\phi_1-\varphi)}\,.
\end{align}
\end{subequations}
Note that the upper limit of the $k_{t1}$ integral is the renormalisation scale $\mu_\mathrm{R}=p_t$, which translates into an upper limit $1$ on $\xi_1$. In order to perform the angular integrations we introduce the polar variables $(r_1,\theta_1)$ such that
\begin{equation}\label{eq:change}
\eta_1-y=R\,r_1\cos\theta_1\,,\qquad\phi_1-\varphi=R\,r_1\sin\theta_1\,,
\end{equation}
and make a change of variables in the integration such that $\d\eta_1\,\d\phi_1=\frac{1}{2}\,R^2\,\d r_1^2\,\d\theta_1$. One may expand the jet mass $\varrho_1$ defined in eq. \eqref{eq:def} as a series in $R$ as follows
\begin{equation}
\varrho_1=\xi_1\,R^2\,r_1^2+\frac{1}{12}\,\xi_1\,R^4\,r_1^4\cos(2\,\theta_1)+\cdots\,.
\end{equation}
In fact, at single logarithmic accuracy it suffices to keep just the first term in this expansion, and thus we write the step function in eq. \eqref{eq:twn} as $\Theta\left(\xi_1\,R^2\,r_1^2-\rho\right)$. We now perform the integrations for each dipole.
\begin{itemize}
\item The dipole $(ab)$:

The contribution of the in-in dipole $(ab)$ to eq. \eqref{eq:twn} at single logarithmic accuracy may be written as follows
\begin{align}
f_{\B,(ab)}^{(1)}(\rho)&=-\mathcal{C}_{ab}\,\bar{\alpha}_s\,\frac{R^2}{2}\int\frac{\d\xi_1}{\xi_1}\,\d r_1^2\,\frac{\d\theta_1}{2\pi}\,\Theta\left(1-r_1^2\right)\times\notag\\
&\quad\times\Theta\left[\xi_1\,R^2\,r_1^2-\rho\right]\notag\\
&=-\mathcal{C}_{ab}\,\bar{\alpha}_s\,\frac{R^2}{2}\,L\,,
\end{align}
with $L=\ln(R^2/\rho)$ being the large logarithm that we aim to resum. This contribution corresponds to soft wide-angle radiation from the in-in dipole into the interior of the measured outgoing jet, and is thus free from collinear logarithms.
\item The dipole $(aj)$:

For the in-jet dipole $(aj)$ eq. \eqref{eq:twn} reads
\begin{align}\label{eq:dSigma_1loop_aj}
&f_{\B,(aj)}^{(1)}=-\mathcal{C}_{aj}\,\bar{\alpha}_s\,\frac{R^2}{2}\int\frac{\d\xi_1}{\xi_1}\,\d r_1^2\,\frac{\d\theta_1}{2\pi}\,\Theta\left[\xi_1\,R^2\,r_1^2-\rho\right]\times\notag\\
&\times\frac{1}{2}\,\frac{\exp(R\,r_1\cos\theta_1)}{\cosh(R\,r_1\cos\theta_1)-\cos(R\,r_1\sin\theta_1)}\,\Theta\left(1-r_1^2\right).
\end{align}
Note here that the step function $\Theta\left[\xi_1\,R^2\,r_1^2-\rho\right]$ which restricts $\xi_1>\rho/(R^2\,r_1^2)$ also implies that $R^2\,r_1^2>\rho$ since $\xi_1<1$. This serves as a collinear regulator for the integral over $r_1$, which would otherwise diverge, resulting in an overall double logarithm as well as a single logarithm. Evaluating the $\xi_1$ integration yields
\begin{align}
&f_{\B,(aj)}^{(1)}(\rho)=-\mathcal{C}_{aj}\,\bar{\alpha}_s\,\frac{R^2}{2}\int_{\rho/R^2}^1\ln\frac{R^2\,r_1^2}{\rho}\,\d r_1^2\,\times\notag\\ &\times\int_0^{2\pi}\frac{\d\theta_1}{2\pi}\,\frac{1}{2}\,\frac{\exp(R\,r_1\cos\theta_1)}{\cosh(R\,r_1\cos\theta_1)-\cos(R\,r_1\sin\theta_1)}\,.
\end{align}
We perform the integration over $\theta_1$ by expanding the integrand as a series in $R$ and neglecting higher-order terms that have small coefficients. Thus we find
\begin{align}
&\frac{R^2}{2}\int_0^{2\pi}\frac{\d\theta_1}{2\pi}\,\frac{1}{2}\,\frac{\exp(R\,r_1\cos\theta_1)}{\cosh(R\,r_1\cos\theta_1)-\cos(R\,r_1\sin\theta_1)}=\notag\\
&=\frac{1}{2\,r_1^2}+\frac{R^2}{8}+\frac{r_1^2\,R^4}{288}+\mathcal{O}(R^8)\,.\label{eq:hg}
\end{align}
The first term in this expansion corresponds to soft {\it and} collinear emissions from the outgoing hard leg $(j)$ into its own jet. It contributes at the double logarithmic level, giving the result
\begin{equation}\label{eq:DLaj}
f_{\B,(aj)}^{(1)\mathrm{DL}}=-\mathcal{C}_{aj}\,\bar{\alpha}_s\,\frac{L^2}{4}\,,
\end{equation}
which is independent of the jet radius (other than in the argument of the logarithm). The other terms in the expansion \eqref{eq:hg} are purely soft wide-angle contributions, hence we can set $\rho\to0$ in the lower limit of integration over $r_1^2$, and throw away the sub-leading $\ln r_1^2$ term in the integrand. Performing the integration we obtain
\begin{equation}\label{eq:SLaj}
f_{\B,(aj)}^{(1)\mathrm{SL}}=-\mathcal{C}_{aj}\,\bar{\alpha}_s\,L\left(\frac{1}{8}\,R^2+\frac{1}{576}\,R^4+\mathcal{O}(R^8)\right).
\end{equation}
We note that the coefficient of $R^8$ in this expression is vanishingly small ($\mathcal{O}(10^{-7})$).
\item The dipole $(bj)$:

For the other in-jet dipole $(bj)$ the only differences relative to the dipole $(aj)$ are the colour factor $\mathcal{C}_{aj}\to\mathcal{C}_{bj}$ and a minus sign to be inserted in the exponent of the exponential in the integrand of eq. \eqref{eq:dSigma_1loop_aj}, i.e., $\exp(R\,r_1\cos\theta_1)\to\exp(-R\,r_1\cos\theta_1)$. This is equivalent to a change $R\to-R$ (the rest of the integral is invariant under this change). This actually does not produce any differences in the integration since only even powers of $R$ appear in the results \eqref{eq:DLaj} and \eqref{eq:SLaj}.
\end{itemize}

We can therefore write the assembled soft-collinear double-logarithmic result as
\begin{equation}
f_{\B,\delta}^{(1)\mathrm{DL}}=-\left(\mathcal{C}_{aj}+\mathcal{C}_{bj}\right)\bar{\alpha}_s\,\frac{L^2}{4}\,,
\end{equation}
and the soft wide-angle single-logarithmic contribution as
\begin{equation}
f_{\B,\delta}^{(1)\mathrm{SL}}=-\bar{\alpha}_s\,L\left[\mathcal{C}_{ab}\,\frac{R^2}{2}+\left(\mathcal{C}_{aj}+\mathcal{C}_{bj}\right)h(R)\right],
\end{equation}
with
\begin{equation}\label{eq:fR}
h(R)=\frac{1}{8}\,R^2+\frac{1}{576}\,R^4+\mathcal{O}(R^8)\,.
\end{equation}
This result was first derived in ref. \cite{Dasgupta:2012hg}, and it actually exponentiates to all orders. However, the running coupling, whose argument is the invariant transverse momentum $\kappa_{t1,(i\ell)}^2=k_{t1}^2/w_{i\ell}^1$ of the emission $k_1$ with respect to the emitting dipole $(i\ell)$ \cite{Catani:1999ss}, contributes at higher orders and modifies the single logarithmic contribution $f_{\B,\delta}^{(1)\mathrm{SL}}$ with a change
\begin{equation}
-\bar{\alpha}_s\,L\to\frac{1}{2\pi\beta_0}\ln\left(1-2\,\alpha_s\,\beta_0\,L\right),
\end{equation}
where $\beta_0$ is the one-loop coefficient of the QCD $\beta$ function. Accounting for the running coupling for the double logarithmic contribution $f_{\B,\delta}^{(1)\mathrm{DL}}$ is more subtle. In fact the running coupling introduces additional single logarithmic components which depend on the renormalisation scheme. We discuss the all-orders resummed result in the following subsection.

\subsection{Resummed global result}

The full NLL-resummed global form factor $f_{\B,\delta}^{\mathrm{global}}(\rho)$ has been computed in ref. \cite{Dasgupta:2012hg} (eqs. (3.3), (3.11) and appendix C therein). The interested reader is referred to the latter reference, together with ref. \cite{Banfi:2004yd}, for details. Here we only state its form, which is given by \cite{Dasgupta:2012hg}
\begin{align}\label{eq:ResummedFormFactor_global}
f_{\B,\delta}^{\mathrm{global}}(\rho)=\frac{1}{\Gamma\left[1+\mathcal{R}'_\delta(\rho)\right]}\exp\left[-\mathcal{R}_\delta(\rho)-\gamma_E\,\mathcal{R}'(\rho)\right],
\end{align}
with $\gamma_E$ the Euler-Mascheroni constant ($\gamma_E\approx0.577$) and $\Gamma$ denotes the Gamma function. The radiator $\mathcal{R}$ and its derivative with respect to $L$, $\mathcal{R}'$, are presented in appendix \ref{app:GlobalFormFactor}. We note that the global form factor is identical for all jet algorithms. We also note that the expression of $f_{\B,\delta}^{\mathrm{global}}(\rho)$ may be deduced from the general form presented in ref. \cite{Banfi:2004yd} as we show in appendix \ref{app:GlobalFormFactor}.

In the next section we treat the case of two-gluon emission where clustering and non-global logarithms first pop-up.

\section{Two-gluon emission}

In the eikonal approximation, the factorised squared amplitude for the emission of two real gluons $k_1$ and $k_2$ off the three-hard-legs Born configuration is given by \cite{Khelifa-Kerfa:2020nlc}
\begin{equation}
\cW_{12,\delta}^{\R\R}=\cW_{1,\delta}^\R\,\cW_{2,\delta}^\R+\cWb_{12,\delta}^{\R\R}\,,
\end{equation}
where the one-loop amplitude squared $\cW_{i,\delta}^{\R}$, which builds up the {\it reducible} part of the above two-gluon squared amplitude (first term on the right-hand side), is given in eq. \eqref{eq:W1R}, and the {\it irreducible} contribution $\cWb_{12,\delta}^{\R\R}$ reads
\begin{equation}
\cWb_{12,\delta}^{\R\R}=\CA\sum_{(i\ell)\in\Delta_\delta}\mathcal{C}_{i\ell}\,\A_{i\ell}^{12}\,.
\end{equation}
The virtual corrections at this order are
\begin{subequations}
\begin{align}
\cW_{12,\delta}^{\R\V}&=-\cW_{12,\delta}^{\R\R}\,,\\
\cW_{12,\delta}^{\V\R}&=-\cW_{1,\delta}^{\R}\,\cW_{2,\delta}^{\R}\,,\\
\cW_{12,\delta}^{\V\V}&=-\cW_{12,\delta}^{\V\R}\,.
\end{align}
\end{subequations}
Following ref. \cite{Khelifa-Kerfa:2015mma}, and implementing the measurement-operator method, we write the jet mass distribution at this order as
\begin{align}\label{eq:dSigma_2loop}
f_{\B,\delta}^{(2)}(\rho)&=\int_{\xi_1>\xi_2}\d\Pi_{12}\,\Xi^{\mathrm{p}}(k_1,k_2)\,\cW_{1,\delta}^\R\,\cW_{2,\delta}^\R-\notag\\
&-\int_{\xi_1>\xi_2}\d\Pi_{12}\,\Xi^{\NG}(k_1,k_2)\,\cWb_{12,\delta}^{\R\R}\,,
\end{align}
with phase-space factor $\d\Pi_{12}=\d\Phi_1\,\d\Phi_2\,\Theta(\varrho_1-\rho)\,\Theta(\varrho_2-\rho)$. Here, the first integral produces the primary-emission contribution, which contains CLs, and the second integral gives NGLs. The functions $\Xi^{\mathrm{p}}$, where $\mathrm{p}$ stands for primary, and $\Xi^{\NG}$, where $\NG$ stands for non-global, result from the application of the jet algorithm and restrict the angular integration regions for gluons $k_1$ and $k_2$.

\subsection{Clustering logarithms}

In this subsection we focus on the primary-emission integral in eq. \eqref{eq:dSigma_2loop} and leave the treatment of the correlated-emission NGLs term to the next subsection. The primary-emission contribution may be split into two parts. The first is the global component which results from integrating both gluons within the measured jet region. This has, however, been accounted for by the all-orders resummed formula \eqref{eq:ResummedFormFactor_global} discussed in the previous section, and will thus be skipped here. The second part is related to the way jet algorithms cluster gluons and results in large single logarithms that are referred to as clustering logarithms \cite{Banfi:2005gj, Delenda:2006nf, Delenda:2012mm}. These logarithms are a result of miscancellation between real emissions and virtual corrections. The key point is that while real gluons may be dragged into/out of the jet by other real gluons and thus get clustered together, virtual gluons can neither drag nor get dragged. We note that CLs are totally absent when jets are reconstructed using the anti-$k_t$ algorithm. At two loops, the C-A and $k_t$ algorithms produce identical CLs, but they start to differ at higher orders as was shown in ref. \cite{Delenda:2012mm}.

To perform the first integral in eq. \eqref{eq:dSigma_2loop} we begin by simplifying the clustering function $\Xi^{\mathrm{p}}(k_1,k_2)$. To this end, we introduce the same change of variables as in eq. \eqref{eq:change} such that $(\eta_1,\phi_1)\to(r_1,\theta_1)$ and $(\eta_2,\phi_2)\to(r_2,\theta_2)$. Note that the upper limit of $r_i$ is $\pi/\left(R\,|\sin\theta_i|\right)$ since we have $-\pi<\phi_i-\varphi<\pi$. We then have for the $k_t$ clustering algorithm \cite{Delenda:2006nf,Delenda:2012mm}
\begin{align}
&\Xi^{\mathrm{p}}(k_1,k_2)=\Theta\left(R^2-d_{1j}\right)\Theta\left(R^2-d_{2j}\right)+\notag\\
&+\Theta\left(d_{1j}-R^2\right)\Theta\left(R^2-d_{2j}\right)\Theta\left(d_{2j}-d_{12}\right)\notag\\
&=\Theta\left(1-r_1^2\right)\Theta\left(1-r_2^2\right)+\notag\\
&+\Theta\left(r_1^2-1\right)\Theta\left(1-r_2^2\right)\Theta\left(2\,r_2\cos(\theta_1-\theta_2)-r_1\right),
\end{align}
where the algorithm distances $d_{im}$ have been defined in eq. \eqref{eq:AlgDistances}. The first term exactly reproduces half the square of the one-loop result \eqref{eq:twn}, i.e.
$1/2!\,[f_{\B,\delta}^{(1)}]^2$, and persists at higher orders as $1/n!\,[f_{\B,\delta}^{(1)}]^n$. This signifies that the one-loop result simply exponentiates into the global form factor discussed before. It is the second term, the CLs term, that we shall focus on in the remainder of this subsection.

We write the CLs contribution at this order as follows
\begin{align}\label{eq:C2_General}
&\cC_{2,\delta}(\rho)=\bar{\alpha}_s^2\,R^4\int_{\xi_1>\xi_2}\frac{\d\xi_1}{\xi_1}\,r_1\,\d r_1\,\frac{\d\theta_1}{2\pi}\,\frac{\d\xi_2}{\xi_2}\,r_2\,\d r_2\,\frac{\d\theta_2}{2\pi}\,\times\notag\\ &\times\Theta\left(r_1^2-1\right)\Theta\left(1-r_2^2\right)\Theta\left(2\,r_2\cos(\theta_1-\theta_2)-r_1\right)\times\notag\\
&\Theta(\varrho_1-\rho)\Theta(\varrho_2-\rho)\left[\sum_{(ik)\in\Delta_\delta}\cC_{ik}\,w_{ik}^1\right]\left[\sum_{(\ell m)\in\Delta_\delta}\cC_{\ell m}w_{\ell m}^2\right].
\end{align}
The expressions inside the square brackets are the one-loop eikonal amplitudes squared \eqref{eq:W1R} for gluons $k_1$ and $k_2$, respectively. To single logarithmic accuracy the $\xi$ integrations factor out from the rest of the integrals yielding the result $L^2/2!$, and we are left with
\begin{align}\label{eq:C2_Channel1_final}
\cC_{2,\delta}(\rho)=\frac{1}{2!}\,\bar{\alpha}_s^2\,L^2\,\mathcal{F}_2^{\delta}(R)\,,
\end{align}
where
\begin{align}
&\mathcal{F}_2^{\delta}(R)=\sum_{(ik)\in\Delta_\delta}\sum_{(\ell m)\in\Delta_\delta}\cC_{ik}\,\cC_{\ell m}\,R^4\int_0^1r_2\,\d r_2\,\frac{\d\theta_2}{2\pi}\,\times\notag\\
&\times\int_1^2r_1\,\d r_1\,\frac{\d\theta_1}{2\pi}\,\Theta\left(2\,r_2\cos(\theta_1-\theta_2)-r_1\right)\,w_{ik}^1\,w_{\ell\,m}^2\notag\\
&=\sum_{(ik)\in\Delta_\delta}\cC_{ik}^2\,\cF_{2,\mathrm{dip}}^{(ik)}+\sum_{(ik)\neq(\ell m)\in\Delta_\delta}\cC_{ik}\,\cC_{\ell m}\,\cF_{2,\mathrm{int}}^{(ik,\ell m)}\,,\label{eq:F2=Fdip+Fint}
\end{align}
where the first term represents contributions from {\it independent} dipoles, that is, each dipole consecutively emits softer gluons at each order independently of the other dipoles. This situation is analogous to that in $e^+e^-$ annihilation to di-jet process (see for instance ref. \cite{Delenda:2012mm}). The second term in eq. \eqref{eq:F2=Fdip+Fint} represents contributions arising from the interference of dipoles in channel $\delta$.

To carry out the integrations we expand the integrand as a power series in $R$ and use the change of variable $\theta_1-\theta_2\to\theta_1$ for the angular integrations. We obtain the following result
\begin{subequations}\label{eq:F2_Dip-Int_coeffs}
\begin{align}
\cF_{2,\mathrm{dip}}^{(ab)}&=0.052\,R^4\,,\\
\cF_{2,\mathrm{dip}}^{(aj)}=\cF_{2,\mathrm{dip}}^{(bj)}&=0.046+0.047\,R^2+0.009\,R^4+\notag\\
&+0.0004\,R^6+\cO(R^8)\,,
\end{align}
for the independent-dipoles part, and
\begin{align}
\cF_{2,\mathrm{int}}^{(aj,bj)}=\cF_{2,\mathrm{int}}^{(bj,aj)}&=0.046+0.004\,R^2+0.0004\,R^4+\notag\\
&+0.00004\,R^6+\cO(R^8)\,,\\
\cF_{2,\mathrm{int}}^{(aj,ab)}=\cF_{2,\mathrm{int}}^{(bj,ab)}&=0.032\,R^2+0.013\,R^4+\notag\\
&+0.0006\,R^6+\cO(R^8)\,,\\
\cF_{2,\mathrm{int}}^{(ab,aj)}=\cF_{2,\mathrm{int}}^{(ab,bj)}&=0.071\,R^2+0.013\,R^4+\notag\\
&+0.0003\,R^6+\cO(R^8)\,,
\end{align}
\end{subequations}
for the dipole-interference part. Notice that the interference term $\cF_{2,\mathrm{int}}$ is not symmetric under the interchange of the dipoles $(aj)$ and $(ab)$, or $(bj)$ and $(ab)$, as apposed to the dipoles $(aj)$ and $(bj)$. This stems from the fact that integrands such as $w_{aj}^1\,w_{ab}^2$ and $w_{aj}^2\,w_{ab}^1$ are not identical, though symmetric under $(r_1,\theta_1)\leftrightarrow(r_2,\theta_2)$. Since the angular restrictions on $k_1$ and $k_2$ are not identical then the results one obtains for the two mentioned terms are different. This boils down to the effect of the $k_t$ algorithm which does not treat the two gluons symmetrically. Furthermore, independent and interference terms involving the in-in $(ab)$ dipole vanish in the limit $R\to0$.

\begin{figure}[ht]
\centering
\resizebox{0.45\textwidth}{!}{\includegraphics{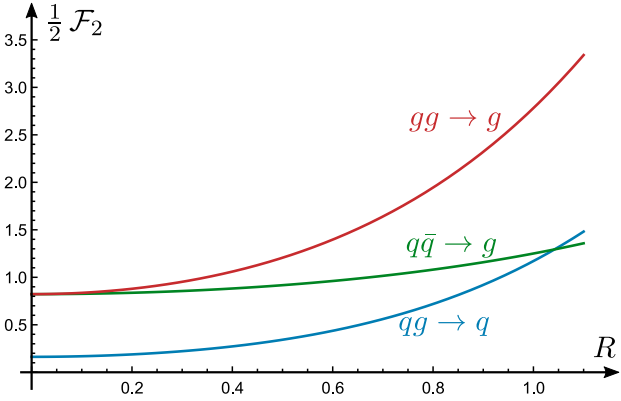}}
\caption{\label{fig:F2}Two-loops CLs coefficient as a function of jet radius $R$ in the $k_t$ and C-A algorithms for the three channels.}
\end{figure}

Substituting the results \eqref{eq:F2_Dip-Int_coeffs} back into eq. \eqref{eq:F2=Fdip+Fint} we obtain the corresponding CLs coefficients for each channel. They read
\begin{subequations}\label{eq:ss}
\begin{align}
&\mathcal{F}_2^{\delta_1}(R)=\CFsq\,0.207\,R^4+\notag\\
&+\CF\,\CA\left[0.414\,R^2-0.103\,R^4+0.004\,R^6+\mathcal{O}(R^{10})\right]+\notag\\
&+\CAsq\left[0.183-0.103\,R^2+0.019\,R^4-0.001\,R^6+\mathcal{O}(R^8)\right],\label{eq:gh}
\end{align}
for channel $q\bar{q}\to g+X$,
\begin{align}
&\mathcal{F}_2^{\delta_2}(R)=\notag\\
&=\CAsq\left[0.087\,R^2+0.069\,R^4+0.001\,R^6+\mathcal{O}(R^{10})\right]+\notag\\
&+\CF\,\CA\left[0.034\,R^2+0.017\,R^4+0.0003\,R^6+\mathcal{O}(R^{10})\right]+\notag\\
&+\CFsq\left[0.183+0.190\,R^2+0.037\,R^4+0.002\,R^6+\mathcal{O}(R^8)\right],\label{eq:qh}
\end{align}
for channel $qg\to q+X$, and
\begin{align}
&\mathcal{F}_2^{\delta_3}(R)=\notag\\
&\CAsq\left[0.183+0.310\,R^2+0.122\,R^4+0.003\,R^6+\mathcal{O}(R^8)\right],
\end{align}
\end{subequations}
for $gg\to g+H$. We show in figure \ref{fig:F2} a plot of the CLs coefficient $\frac{1}{2!}\,\mathcal{F}_2^{\delta}$ as a function of $R$ for the various channels $\delta$.
We notice that gluon-initiated jets have larger CLs coefficient than quark-initiated jets, mainly due to the corresponding colour factors ($\CA=3$ and $\CF=4/3$, respectively). These series expansions in $R$ converge, and at small values of $R$ it suffices to keep only the leading terms. At very small values of $R$ we observe that
\begin{equation}\label{eq:F2_LimitR=0}
\lim_{R\to0}\mathcal{F}_2^{\delta_{1,3}}=\CAsq\,0.183\,,\qquad
\lim_{R\to0}\mathcal{F}_2^{\delta_{2}}=\CFsq\,0.183\,.
\end{equation}
This result for CLs obtained here for $H/W/Z/\gamma$ + jet events at hadron colliders coincides with that found in refs. \cite{Delenda:2012mm, Banfi:2010pa, KhelifaKerfa:2011zu} for jet mass distribution in $e^+e^-\to$ di-jet events. It does, however, deviate from it as $R$ increases due to initial-state radiation from the incoming partons. Inline with the findings of refs. \cite{Delenda:2006nf, Delenda:2012mm} we expect the term \eqref{eq:C2_Channel1_final} to simply exponentiate to all orders. Nonetheless, there will be new CLs terms at each order that are not captured by the latter exponential and that are highly non-trivial to deduce (see ref. \cite{Delenda:2012mm}). Moreover, we expect that at higher orders the small-$R$ limit of the CLs coefficient in $H/V$ + jet events at the LHC will coincide with that in $e^+e^-\to$ di-jet events found in ref. \cite{Delenda:2012mm}.

In the next subsection we compute NGLs at two loops.

\subsection{Non-global logarithms}

\subsubsection{k$_\text{t}$ and C-A clustering algorithms}
\label{subsec:2-loop_kt-CA}

\begin{figure}[ht]
\centering
\resizebox{0.45\textwidth}{!}{\includegraphics{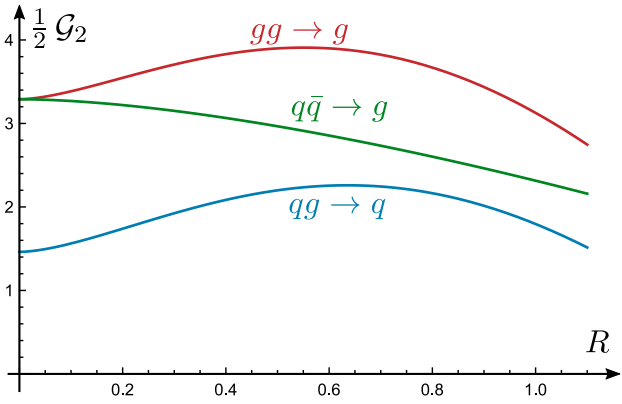}}
\caption{\label{fig:G2}Two-loops NGLs coefficient as a function of jet radius for C-A and $k_t$ clustering.}
\end{figure}

We now turn to the evaluation of the correlated secondary-emission contribution in eq. \eqref{eq:dSigma_2loop} for the $k_t$ and C-A clustering algorithms. To this end we write
\begin{align}
\mathcal{S}_{2,\delta}(\rho)&=-\int_{\xi_1>\xi_2}\d\Phi_1\,\d\Phi_2\,\Theta\left(\varrho_1-\rho\right)\Theta\left(\varrho_2-\rho\right)\times\notag\\
&\times\Xi^{\NG}(k_1,k_2)\left[\sum_{(i\ell)\in\Delta_\delta}\CA\,\mathcal{C}_{i\ell}\,\A_{i\ell}^{12}\right],\label{eq:S2}
\end{align}
where the clustering function reads
\begin{align}
&\Xi^{\NG}(k_1,k_2)=\Theta\left(d_{1j}-R^2\right)\Theta\left(R^2-d_{2j}\right)\Theta\left(d_{12}-d_{2j}\right)\notag\\
&=\Theta\left(r_1^2-1\right)\Theta\left(1-r_2^2\right)\Theta\left(r_1-2\,r_2\cos(\theta_1-\theta_2)\right).
\end{align}
As before, the integration over $\xi_1$ and $\xi_2$ yields $L^2/2!$, and we may write
\begin{subequations}
\begin{align}
\mathcal{S}_{2,\delta}(\rho)&=-\frac{1}{2!}\,\bar{\alpha}_s^2\,L^2\,\mathcal{G}_{2}^{\delta}(R)\,,
\end{align}
with NGLs coefficient
\begin{align}\label{eq:G2_kt}
&\mathcal{G}_2^{\delta}(R)=\CA\sum_{(i\ell)\in\Delta_\delta}\cC_{i\ell}\,R^4\int_0^1 r_2\,\d r_2\,\frac{\d\theta_2}{2\pi}\,\times\notag\\
&\times\int_1^{\pi/\left(R\,|\sin\theta_1|\right)}r_1\,\d r_1\,\frac{\d\theta_1}{2\pi}\,\Theta\left(r_1-2\,r_2\cos(\theta_1-\theta_2)\right)\A_{i\ell}^{12}\notag\\
&=\CA\sum_{(i\ell)\in\Delta_\delta}\cC_{i\ell}\,\mathcal{G}_{2}^{(i\ell)}(R)\,.
\end{align}
\end{subequations}
Performing the integration, as in the previous subsection, we obtain the results for $\mathcal{G}_{2}^{(i\ell)}$ for each dipole as a series in $R$
\begin{subequations}
\begin{align}
\mathcal{G}^{(ab)}_2(R)&=-R^2\ln R+0.015\,R^2+0.151\,R^4-\notag\\
&-0.004\,R^6+\mathcal{O}(R^8)\,,\\
\mathcal{G}^{(aj)}_2(R)&=\mathcal{G}^{(bj)}_2(R)=0.366-0.103\,R^2+0.004\,R^4+\notag\\
&+0.0002\,R^6+\mathcal{O}(R^8)\,.
\end{align}
\end{subequations}
In terms of channels we have
\begin{subequations}\label{eq:NGLs_kt_Channels}
\begin{align}
&\mathcal{G}^{\delta_1}_2=\CF\,\CA\left[-2\,R^2\ln R+0.031\,R^2+0.302\,R^4-\right.\notag\\
&\left.-0.008\,R^6+\mathcal{O}(R^{8})\right]+\CAsq\left[0.731+R^2\ln R-0.222\,R^2-\right.\notag\\
&\left.-0.143\,R^4+0.004\,R^6+\mathcal{O}(R^{8})\right],
\end{align}
for channel $q\bar{q}\to g+X$,
\begin{align}
&\mathcal{G}^{\delta_2}_2=\CF\,\CA\left[0.731-0.207\,R^2+0.008\,R^4+0.0004\,R^6+\right.\notag\\
&\left.\mathcal{O}(R^8)\right]+\CAsq\left[-R^2\ln R+0.015\,R^2+0.151\,R^4-\right.\notag\\
&\left.-0.004\,R^6+\mathcal{O}(R^8)\right],
\end{align}
for channel $qg\to q+X$, and
\begin{align}
&\mathcal{G}^{\delta_3}_2=\CAsq\left[0.731-R^2\ln R-0.191\,R^2+0.159\,R^4-\right.\notag\\
&\left.-0.003\,R^6+\mathcal{O}(R^8)\right],
\end{align}
\end{subequations}
for $gg\to g+H$. Moreover, in the small-$R$ limit we observe that
\begin{align}
\lim_{R\to0}\mathcal{G}_2^{\delta_{1,3}}=\CAsq\,0.731\,, \qquad
\lim_{R\to0}\mathcal{G}_2^{\delta_{2}}  =\CF\,\CA\,0.731\,.
\end{align}
The result for channel $\delta_2$ is exactly the same small-$R$ limit found in the case of jet shapes in $e^+e^-\to$ di-jet events (see for instance ref. \cite{KhelifaKerfa:2011zu}). Results for channels $\delta_1$ and $\delta_3$ in the limit $R\to0$ are also the same and differ from those for channel $\delta_2$ only in the colour factor. In figure \ref{fig:G2} we plot the NGLs coefficient $\frac{1}{2!}\,\mathcal{G}_2^{\delta}$ at this order as a function of jet radius $R$.
Once again we notice that gluon-initiated jets have larger NGLs coefficient due to their large gluon-emission colour factor ($\CA$). We observe from the plots in figures \ref{fig:F2} and \ref{fig:G2} that the CLs coefficient for the $gg\to g$ channel grows larger with $R$ while that for NGLs does not change much.

\begin{figure}[ht]
\centering
\resizebox{0.45\textwidth}{!}{\includegraphics{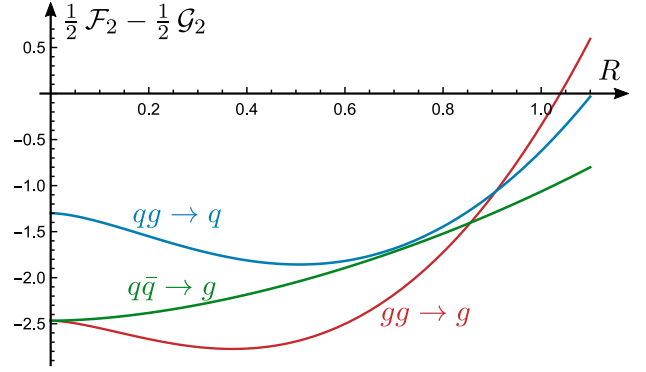}}
\caption{\label{fig:F2vsG2}The coefficient of the single logarithm $\asb^2\,L^2$ resulting from both CLs and NGLs with C-A and $k_t$ clustering.}
\end{figure}

Moreover, in order to assess the overall impact of CLs and NGLs at this order, we plot in figure \ref{fig:F2vsG2} the combined coefficient of the single logarithm $\bar{\alpha}_s^2\,L^2$ resulting from the non-global nature of our observable.
We note that at large jet radii ($R\gtrsim 1.0$) and for all partonic channels the CLs and NGLs tend to balance each other out, but not entirely though. For small jet radii the said single logarithmic CLs + NGLs coefficient is quite large in magnitude especially for gluon-initiated jets.

\subsubsection{Anti-k$_\text{t}$ clustering algorithm}

For the sake of assessing the effect of clustering on NGLs, we report the results for the NGLs coefficient in the anti-$k_t$ algorithm. Note that there are no CLs in this case. The corresponding integral is identical to that in eq. \eqref{eq:S2} except for the clustering function. It reads
\begin{align}
\mathcal{S}^{\mathrm{ak_t}}_{2,\delta}(\rho)&=-\int_{\xi_1>\xi_2}\d\Phi_1\,\d\Phi_2\,\Theta\left(\varrho_1-\rho\right)\Theta\left(\varrho_2-\rho\right)\times\notag\\
&\times\Theta\left(r_1^2-1\right)\Theta\left(1-r_2^2\right)\left[\sum_{(i\ell)\in\Delta_\delta}\CA\,\mathcal{C}_{i\ell}\,\A_{i\ell}^{12}\right]\notag\\
&=-\frac{1}{2!}\,\bar{\alpha}_s^2\,L^2\,\mathcal{G}_{2}^{\delta,\mathrm{ak_t}}(R)\,.\label{eq:2loops}
\end{align}
The results we obtain for each dipole are
\begin{subequations}
\begin{align}
\mathcal{G}^{(ab),\mathrm{ak_t}}_2&=-R^2\ln R+0.500\,R^2+0.125\,R^4-\notag\\
&-0.003\,R^6+\mathcal{O}(R^8)\,,\\
\mathcal{G}^{(aj),\mathrm{ak_t}}_2&=\mathcal{G}^{(bj),\mathrm{ak_t}}_2=0.822+0.003\,R^4+\cO(R^8)\,.
\end{align}
\end{subequations}
In terms of channels we have
\begin{subequations}\label{eq:NGLs_akt_Channels}
\begin{align}
&\mathcal{G}^{\delta_1,\mathrm{ak_t}}_2=\CF\,\CA\left[-2\,R^2\ln R+R^2+0.250\,R^4-\right.\notag\\
&\left.-0.007\,R^6+\mathcal{O}(R^8)\right]+\CAsq\left[1.645+R^2\ln R-0.500\,R^2-\right.\notag\\
&\left.-0.118\,R^4+0.003\,R^6+\mathcal{O}(R^8)\right],\\
&\mathcal{G}^{\delta_2,\mathrm{ak_t}}_2=\CF\,\CA\left[1.645+0.007\,R^4+\mathcal{O}(R^8)\right]+\notag\\
&+\CAsq\left[-R^2\ln R+0.500\,R^2+0.125\,R^4-0.003\,R^6+\right.\notag\\
&\left.+\mathcal{O}(R^8)\right],\\
&\mathcal{G}^{\delta_3,\mathrm{ak_t}}_2=\CAsq\left[1.645-R^2\ln R+0.500\,R^2+0.132\,R^4-\right.\notag\\
&\left.-0.003\,R^6+\mathcal{O}(R^8)\right].
\end{align}
\end{subequations}
These results are in agreement with those reported in ref. \cite{Dasgupta:2012hg}. Notice again that the $R\to0$ limit of the above expressions produces a result (which is proportional to $1.645=\zeta_2$) that is identical to that reported in ref. \cite{KhelifaKerfa:2011zu} for $e^+e^-\to$ di-jet process.

\begin{figure}[ht]
\centering
\resizebox{0.45\textwidth}{!}{\includegraphics{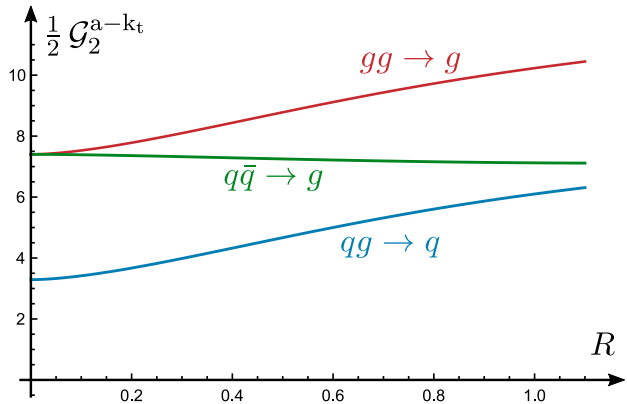}}
\caption{\label{fig:G2gakt}NGLs coefficient with anti-$k_t$--clustered jets at two loops.}
\end{figure}

\begin{figure}[ht]
\centering
\resizebox{0.45\textwidth}{!}{\includegraphics{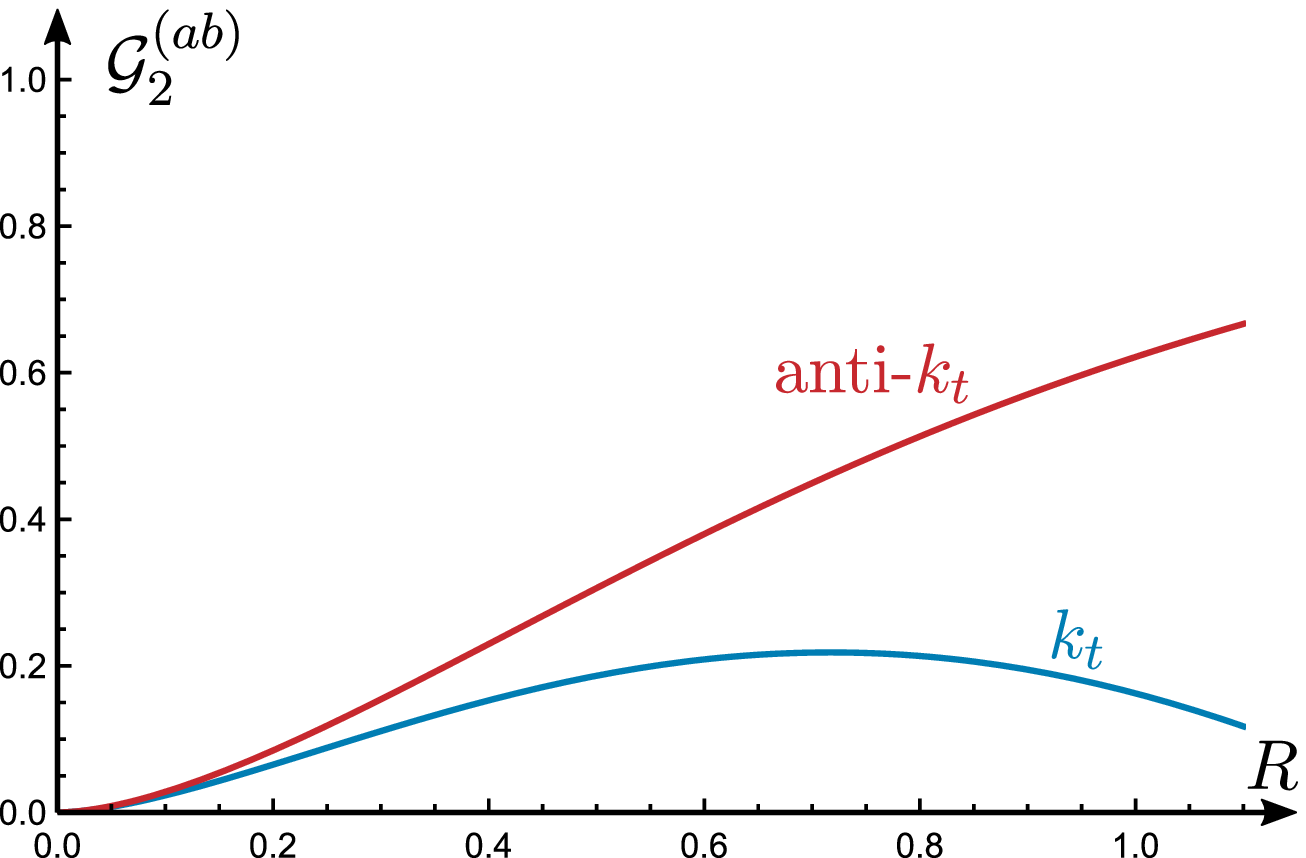}}
\resizebox{0.45\textwidth}{!}{\includegraphics{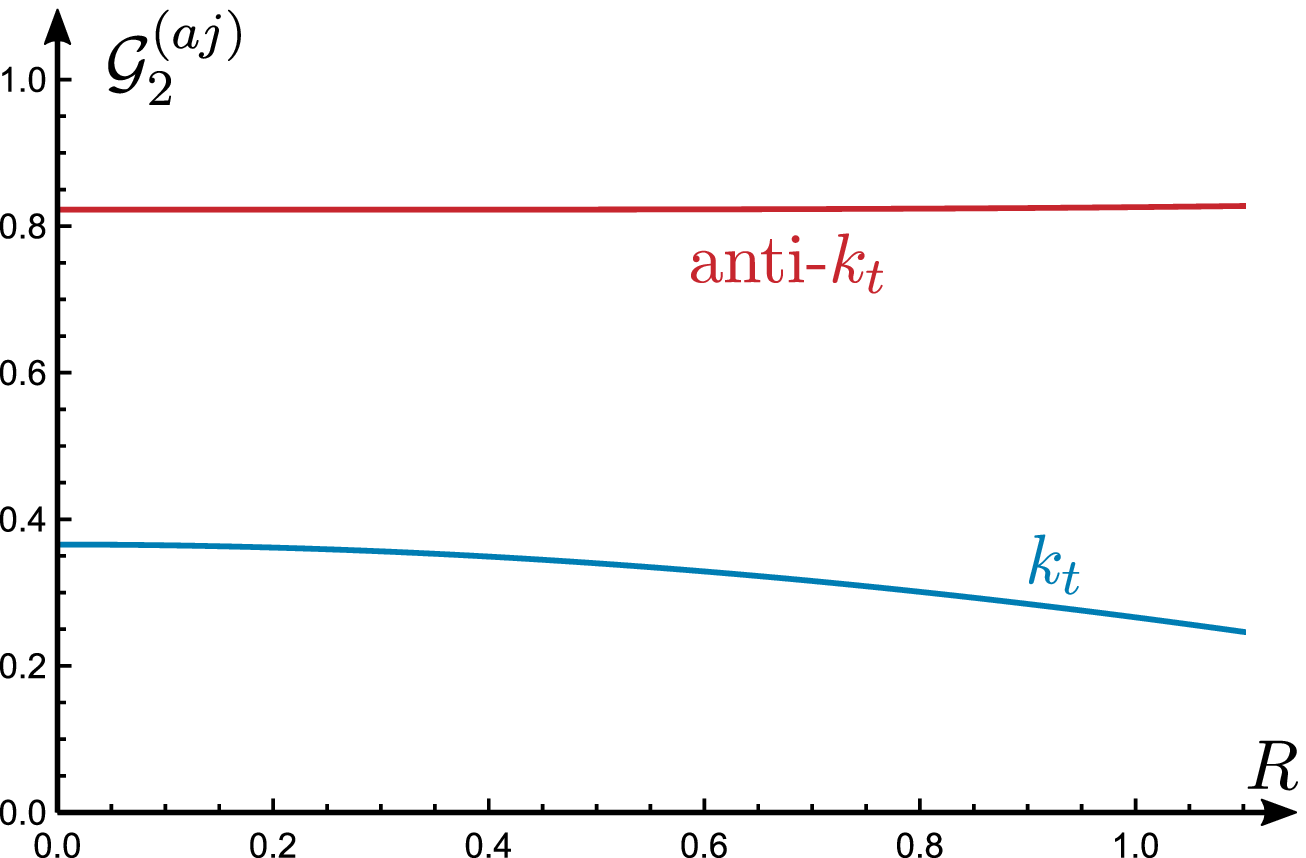}}
\caption{\label{fig:G2_kt-Akt}NGLs coefficients at two loops with anti-$k_t$ and $k_t$ clustering.}
\end{figure}

We plot in figure \ref{fig:G2gakt} the NGLs coefficient $\frac{1}{2!}\,\mathcal{G}_{2}^{\delta,\mathrm{ak_t}}$ with anti-$k_t$--clustered jets as a function of the jet radius $R$ for the various partonic channels.
As is clearly evident from the plots, NGLs in the anti-$k_t$ algorithm are much larger compared to those in the C-A or $k_t$ clustering case. This is made clearer in figure \ref{fig:G2_kt-Akt} where NGLs coefficients for each dipole are plotted for both $k_t$ and anti-$k_t$ algorithms.
This observation was also made in previous studies of NGLs with $k_t$ clustering \cite{Delenda:2006nf,Delenda:2012mm,Appleby:2002ke}. While $k_t$ clustering induces another tower of large single logarithms, namely CLs, it actually diminishes the impact of NGLs. Additionally, as we observed in the previous subsection \ref{subsec:2-loop_kt-CA}, the induced CLs play a role of further reducing NGLs since their coefficients have opposite signs. This may hint at a (phenomenological) favour for the $k_t$ (or C-A) clustering algorithm over the anti-$k_t$ algorithm.

\section{All-orders treatment of CLs and NGLs}

Including the resummation of NGLs and CLs together with the global form factor \eqref{eq:ResummedFormFactor_global} then the all-orders NLL-resummed jet mass distribution may be cast into
\begin{equation}\label{eq:all}
\frac{\d\Sigma_\delta(\rho)}{\d\B_\delta}=\frac{\d\sigma_{0,\delta}}{\d\B_\delta}\,\mathcal{S}_\delta(\rho)\,\mathcal{C}_\delta(\rho)\,f_{\B,\delta}^{\mathrm{global}}(\rho)\,C_{\B,\delta}(\rho)\,,
\end{equation}
where $\mathcal{S}_\delta(\rho)$ and $\mathcal{C}_\delta(\rho)$ account for the resummation of NGLs and CLs, respectively. We note that, unlike the global form factor, the factors $\mathcal{S}_\delta$ and $\mathcal{C}_\delta$ are algorithm-dependent.

In the anti-$k_t$ algorithm, the NGLs form factor results from multiple correlated gluons outside the jet that coherently emit the softest gluon into the jet. For the $k_t$ and C-A clustering algorithms, gluons can be moved into and out of the jet by the clustering, thus NGLs can be induced when more than one gluon is emitted within the jet region from an ensemble of harder gluons. The NGLs factor $\mathcal{S}_\delta$ can be computed numerically and in general only in the large-$\Nc$ limit \cite{Dasgupta:2001sh,Banfi:2002hw}. For the $e^+e^-\to$ di-jet process, finite-$\Nc$ results do exist though \cite{Hatta:2013iba, Hagiwara:2015bia}. Moreover, the CLs form factor results from multiple independent (primary) emissions that are clustered by the $k_t$ or C-A algorithm. Just like NGLs, the latter CLs can also be resummed numerically.

\begin{figure}[ht]
\centering
\resizebox{0.45\textwidth}{!}{\includegraphics{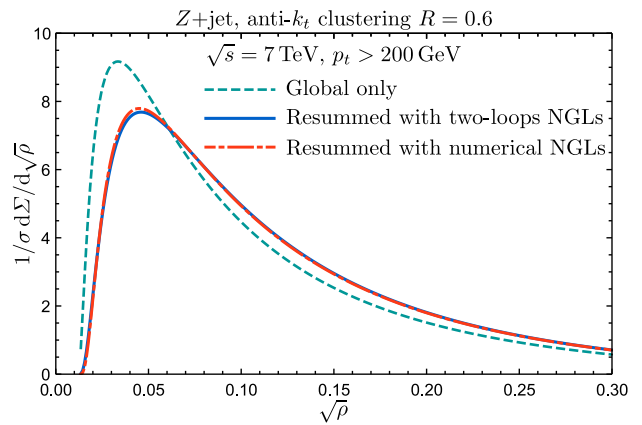}}
\caption{\label{fig:All-vs-Exp2Loop}The full resummed differential jet mass distribution in the anti-$k_t$ jet algorithm with NGLs factor as an exponential of the two-loops result and as an all-orders numerical result obtained from ref. \cite{Dasgupta:2012hg}. We explain the in the next section how these plots are obtained.}
\end{figure}

For the anti-$k_t$ algorithm, the all-orders numerical resummation of NGLs may be obtained from the dipole-evolution Monte Carlo code of ref. \cite{Dasgupta:2001sh} as reported in ref. \cite{Dasgupta:2012hg} for the various dipoles. We see from figure \ref{fig:All-vs-Exp2Loop}
that the exponential of the two-loops result \eqref{eq:2loops} approximates very well the all-orders numerical result for the NGLs factor in the $Z$ + jet process. The same is observed for the other processes. Hence we shall confine ourselves to simply using the exponential of the two-loops result for the $k_t$ and C-A algorithms. To this end we write, for a given channel $\delta$,
\begin{equation}\label{eq:NGLs_ResumFactor}
\mathcal{S}_\delta(\rho)\approx\exp\left[-\frac{1}{2!}\,\mathcal{G}_2^{\delta}(R)\,t^2\right],
\end{equation}
where $\cG_2^\delta$, for the $k_t$ and C-A algorithms, is given in eq. \eqref{eq:NGLs_kt_Channels}, and the evolution parameter $t$ is defined by
\begin{align}\label{eq:EvolParam_t_def}
t=-\frac{1}{2\pi\beta_0}\,\ln\left(1-2\,\alpha_s\,\beta_0\,L\right).
\end{align}
Note that at fixed order $t$ reduces to just $\bar{\alpha}_s\,L$.

As for CLs, it was shown in refs. \cite{Delenda:2006nf, Delenda:2012mm} that the perturbative CLs series exhibits a pattern of exponentiation, and that the exponential of the two-loops result is a very good approximation to the numerically-resummed CLs factor obtained from the code of ref. \cite{Dasgupta:2001sh}. Therefore, and just as we did with NGLs, we shall be using the exponential of the two-loops result for the CLs resummed factor $\mathcal{C}_\delta(\rho)$. Thence
\begin{align}\label{eq:CLs_ResumFactor}
\mathcal{C}_\delta(t)\approx\exp\left[\frac{1}{2!}\,\mathcal{F}_2^{\delta}(R)\, t^2\right],
\end{align}
where $\mathcal{F}_2^\delta$ is given for $k_t$ and C-A algorithms in eq. \eqref{eq:ss}.

\section{Comparison to Pythia 8 and Herwig 7 parton showers}

In this section we present comparisons of our results for the jet mass distribution with those obtained from \texttt{Pythia 8} \cite{Sjostrand:2014zea} and \texttt{Herwig 7} \cite{Bahr:2008pv,Bellm:2015jjp} parton showers (PS), where the jets are clustered with \texttt{FastJet} \cite{Cacciari:2011ma}. The resummed result is obtained by convoluting $\d\Sigma_\delta/\d\B_\delta$ given in eq. \eqref{eq:dSigma} with parton distribution functions (we use MSTW 2008 (NLO) PDFs \cite{Martin:2009iq} and $\mu_\mathrm{F}=\mu_\mathrm{R}=200\,\mathrm{GeV}$). For double-checking we perform the convolution using two different methods. In one method we simply use a Monte Carlo code to integrate over the momentum fractions of the partons $x_a$ and $x_b$ and over the transverse momentum $p_t$ and rapidity $y$ of the jet, as explained in detail in appendix \ref{sec:Born}. In the other approach we generate a set of unweighted parton-level Born events using \texttt{MadEvent} from \texttt{MadGraph} \cite{Maltoni:2002qb, Alwall:2014hca} in the ``\texttt{Les Houches Event File}'' format \cite{Alwall:2006yp}, with the cuts $\Xi_\B$ being applied. We then weigh each event by the resummed form factor $\mathcal{S}_\delta(\rho)\,\mathcal{C}_\delta(\rho)\,f_{\B,\delta}^{\mathrm{global}}(\rho)$, sum over all events, and divide by the effective luminosity $\mathcal{L}=N_{\mathrm{tot}}/\sigma_0$, with $N_{\mathrm{tot}}$ the total number of events and $\sigma_0$ the Born cross-section calculated with \texttt{MadGraph}. This results in the integrated distribution given in eq. \eqref{eq:Integrated_Sogma_delta} from which the differential distribution can straightforwardly be obtained. To avoid low-$p_t$ resummation we impose a cut on $p_t$ of the final-state jet, e.g., $p_t>200\,\mathrm{GeV}$, i.e., we only consider high-$p_t$ jets, at a centre-of-mass energy $\sqrt{s}=7\,\mathrm{TeV}$.

In our resummed result we also include an approximation to the NLO effects on the distribution through the NLO factor $C^{(1)}_{\mathcal{B},\delta}(\rho)$. The full NLO distribution may ideally be analytically calculated using the full squared amplitude with two partons in the final state as well as virtual corrections to the Born cross-section. Though possible this is a delicate task. The alternative numerical approach would be to exploit fixed-order programs and obtain the factor $C^{(1)}_{\mathcal{B},\delta}(\rho)$ as a fully differential distribution in the {\it Born configuration}, and then perform the integration including the resummed form factor over the Born kinematics. Practically this is not feasible. Instead, one could obtain an NLO factor $C^{(1)}_{\mathcal{B},\delta}(\rho)$ that is averaged over the Born configuration \cite{Banfi:2010xy} and insert it in eq. \eqref{eq:dSigma} as if it were unintegrated over $\d\B_\delta$. In this paper we employ this method and estimate the Born-configuration--averaged factor $C^{(1)}_{\delta}(\rho)$ as was done in refs. \cite{Dasgupta:2012hg,Banfi:2010xy}, using the NLO jet mass distribution obtained from the fixed-order program \texttt{MCFM} \cite{Campbell:2015qma,Campbell:2019dru}.

In refs. \cite{Dasgupta:2012hg,Banfi:2010xy}, the NLO factor $C^{(1)}_{\delta}(\rho)$ was calculated in the small-$\rho$ limit as a constant, and then the $\rho$-dependence of the NLO contribution to the jet mass distribution was included at the stage of matching. This is equivalent to using the full $\rho$-dependence of $C^{(1)}_{\delta}(\rho)$, which we do in the present work. The factor $\as\,C^{(1)}_{\delta}(\rho)$ for channel $(\delta)$ is simply given by the NLO integrated jet mass distribution $\Sigma_{\mathrm{NLO}}^{(\delta)}(\rho)$ (obtained from \texttt{MCFM}) minus the expansion of the integrated pure-resummed distribution $\Sigma^{(\delta)}_{\mathrm{NLL},\as}(\rho)$, and then the result is divided by the Born cross-section $\sigma_{0,\delta}$ \cite{Banfi:2010xy}
\begin{equation}
\as\,C^{(1)}_{\delta}(\rho)=\frac{1}{\sigma_{0,\delta}}\left(\Sigma_{\mathrm{NLO}}^{(\delta)}(\rho)-\Sigma^{(\delta)}_{\mathrm{NLL},\as}(\rho)\right).
\end{equation}
At NLO there are new channels that open up, specifically processes with incoming $qq'$ or two gluons, that are not present at the Born level. These channels are not logarithmically enhanced and only contribute a small  correction to the distribution. \footnote{In the current version of \texttt{MCFM}, the only possibility is to separate the channels with incoming $q q'$, $q g$, and $gg$. The former mixes channels with incoming $q\bar{q}$ and the remaining $qq'$ processes, so it is not possible to obtain a clean $q\bar{q}$ channel contribution from this program.}

\begin{figure}[ht]
\centering
\resizebox{0.45\textwidth}{!}{\includegraphics{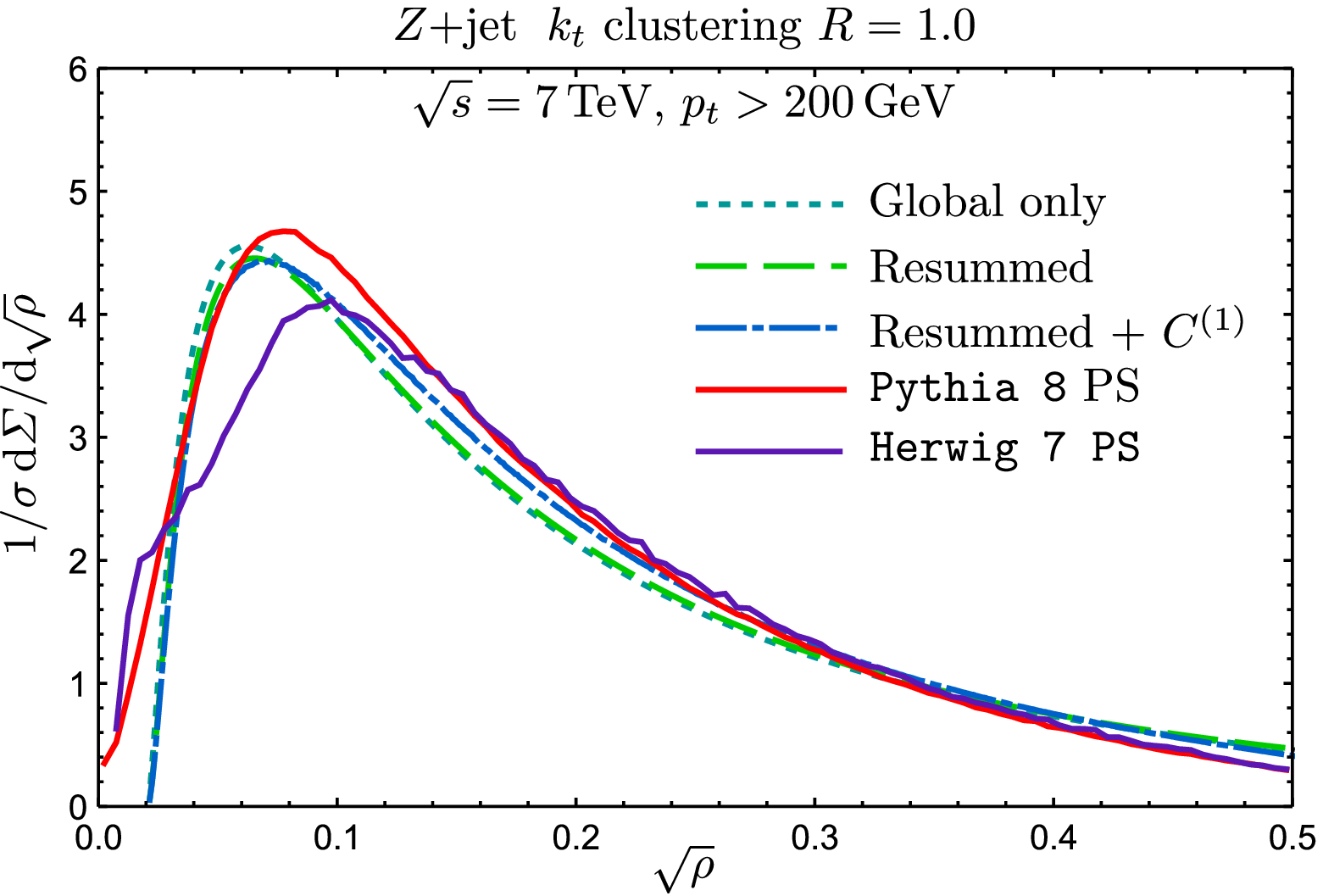}}
\resizebox{0.45\textwidth}{!}{\includegraphics{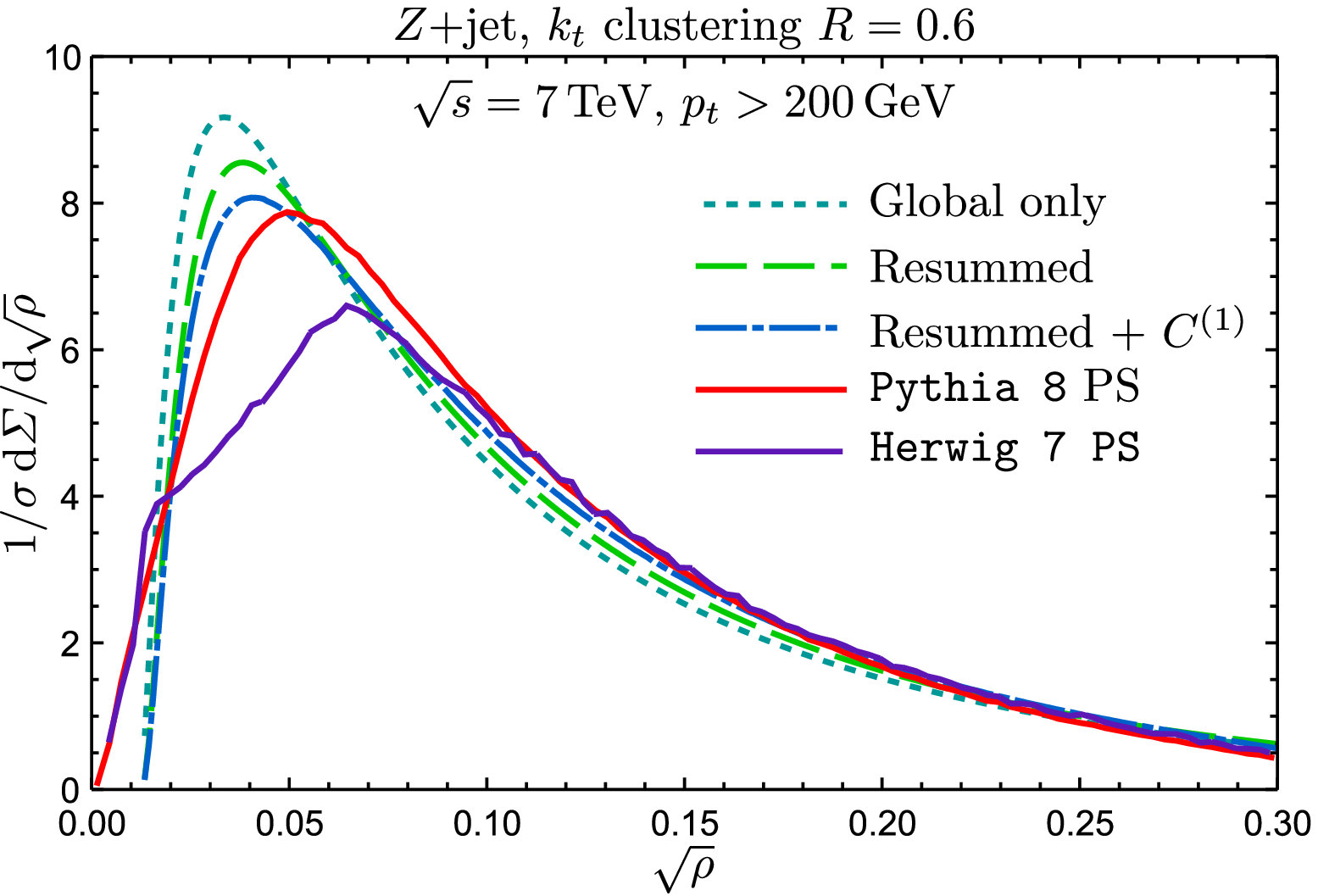}}
\caption{\label{fig:pythiaZ}Resummed differential jet mass distribution in $Z$ + jet events with $k_t$ clustering compared to \texttt{Pythia 8} and \texttt{Herwig 7} PS.}
\end{figure}

In figure \ref{fig:pythiaZ} we show plots for the differential jet mass distribution $1/\sigma\,\d\Sigma/\d\sqrt{\rho}$, where $\Sigma(\rho)$ is defined in eq. \eqref{eq:Integrated_Sogma_delta}, in $Z$ + jet events at the LHC with $k_t$ clustering. We choose two values for the jet radius, one for which the size of NGLs + CLs is expected to be small, $R=1.0$, and another where NGLs + CLs are expected to be important, e.g., $R=0.6$. The global and pure-resummed distributions are normalised to the Born cross-section, while the resummed + $C^{(1)}$, \texttt{Pythia 8}, and \texttt{Herwig 7} distributions are normalised to the total cross-section.

We observe from the $R=1.0$ plot in figure \ref{fig:pythiaZ} that the global and full-resummed distributions are quite close to the \texttt{Pythia 8} PS result, indicating the smallness of the effect of NGLs and CLs factors in this case. For the $R=0.6$ plot, there is a clear difference between the global and \texttt{Pythia 8} PS curves, and our full resummation, which is based on the exponential of the two-loops NGLs and CLs result, seems to do better. We also note that the NLO term $C^{(1)}$ slightly modifies the peak and tail of the distribution especially for $R=0.6$, bringing it even closer to the \texttt{Pythia 8} PS result.

As is clear from the plots, the \texttt{Pythia 8} PS result seems to be in better agreement with our resummed distribution near the peak than \texttt{Herwig 7}. This observation was also made in ref. \cite{Dasgupta:2012hg}. It should be noted, however, that a more comprehensive comparison is feasible only when one includes non-perturbative effects, where different event generators are then expected to be in agreement. We do this in the next section.

\begin{figure}[ht]
\centering
\resizebox{0.45\textwidth}{!}{\includegraphics{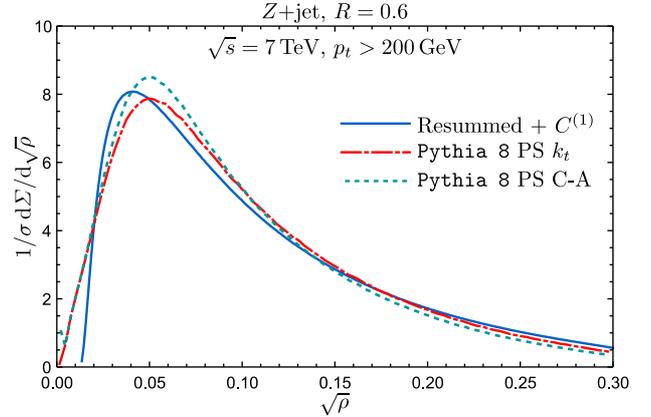}}
\caption{\label{fig:pythiaZCA}Resummed differential jet mass distribution in $Z$ + jet events with C-A algorithm compared to \texttt{Pythia 8} PS.}
\end{figure}

In figure \ref{fig:pythiaZCA} we plot the same distribution employing the C-A algorithm. We recall that up to two loops both $k_t$ and C-A algorithms produce identical results. This means that the resummed formula that includes the exponential of the two-loops NGLs and CLs as well as the $C^{(1)}_{\delta}$ terms for all channels are the same in both algorithms. We expect, however, differences between the two cases when one performs an all-orders NGLs and CLs resummation, and also when one includes the higher-order $C^{(n)}$ terms. We compare, in figure \ref{fig:pythiaZCA}, the resummed + $C^{(1)}$ result with the \texttt{Pythia} 8 PS result employing both $k_t$ and C-A algorithms, where we notice that the peak of the distribution is slightly higher in the latter algorithm.

\begin{figure}[ht]
\centering
\resizebox{0.45\textwidth}{!}{\includegraphics{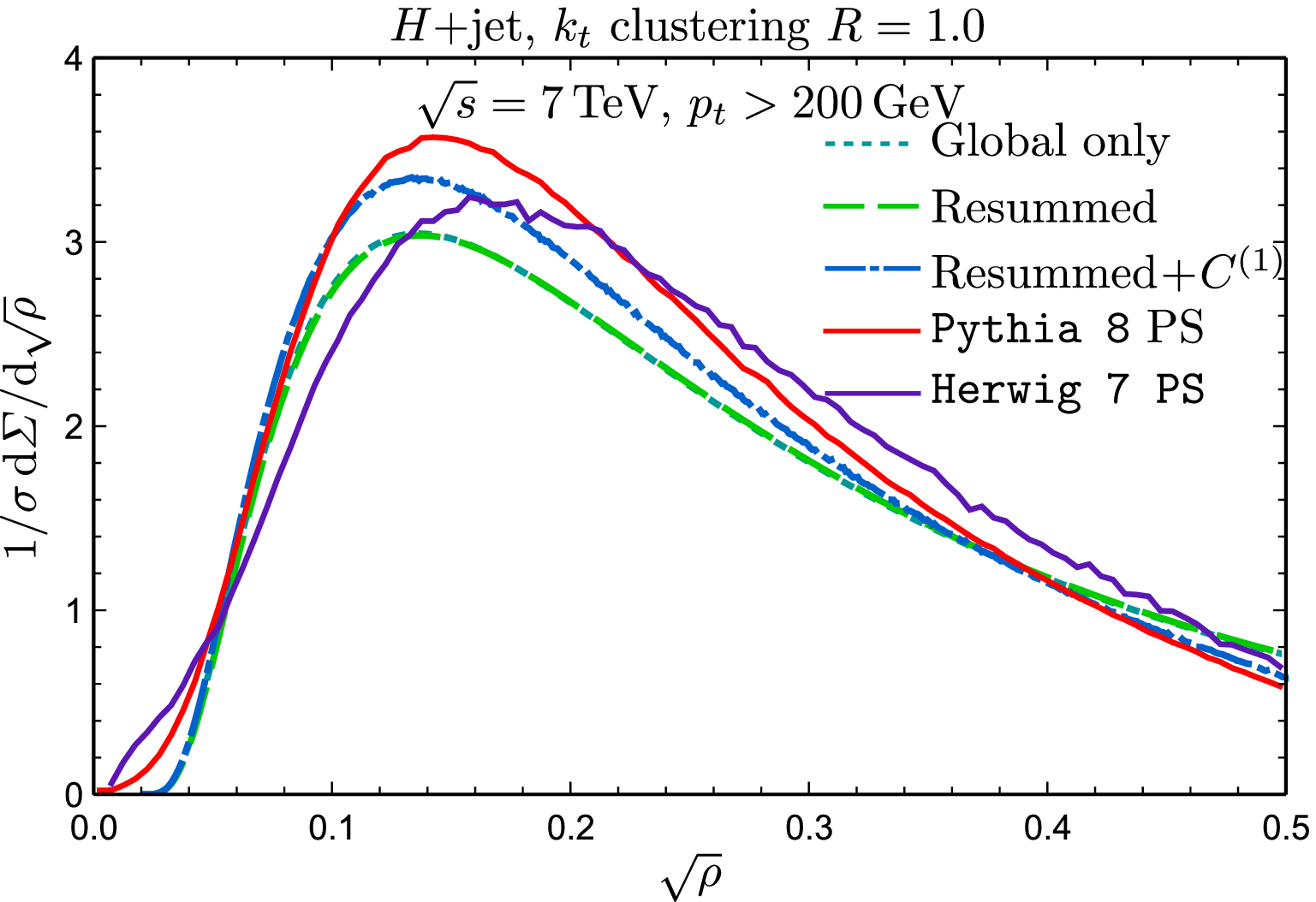}}
\resizebox{0.45\textwidth}{!}{\includegraphics{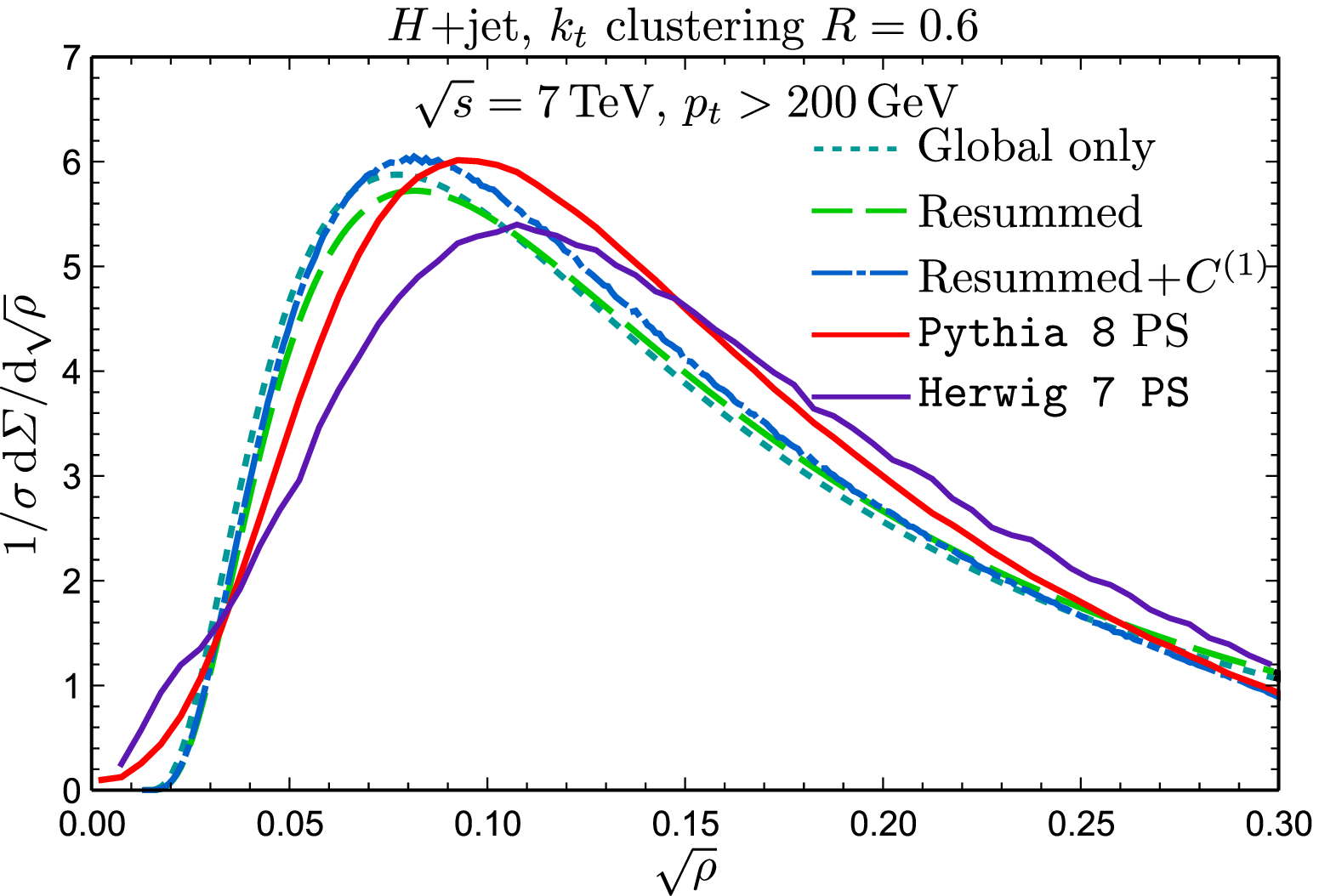}}
\caption{\label{fig:pythiaH}Resummed differential jet mass spectrum with $k_t$ clustering compared to \texttt{Pythia 8} and \texttt{Herwig 7} parton showers for $R=1.0$ and $R=0.6$ in the process $gg\to gH$.}
\end{figure}

We additionally show in figure \ref{fig:pythiaH} the differential jet mass distribution in the process $gg\to Hg$ for jet radii $R=1.0$ and $R=0.6$.
Notice from figures \ref{fig:F2vsG2} and \ref{fig:G2gakt} that, for this channel, the combined effect of NGLs and CLs at two loops in the $k_t$ algorithm is small compared to that of NGLs with anti-$k_t$ clustering, and so we expect our resummed distribution to fit well with the PS result in the case of $k_t$ clustering. This is indeed the case as is clear from figure \ref{fig:pythiaH}, particularly for \texttt{Pythia 8} PS.

\begin{figure}[ht]
\centering
\resizebox{0.45\textwidth}{!}{\includegraphics{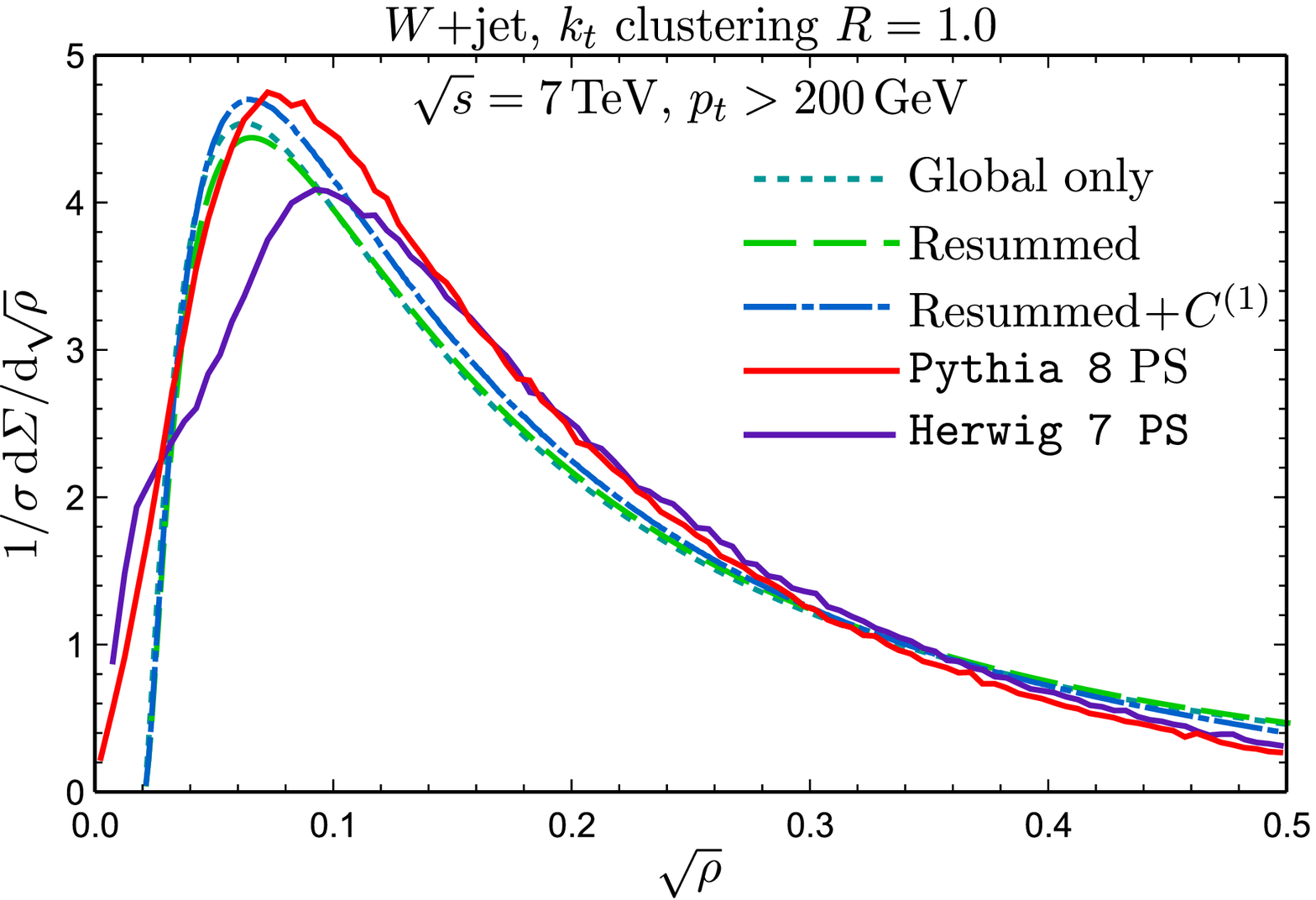}}
\resizebox{0.45\textwidth}{!}{\includegraphics{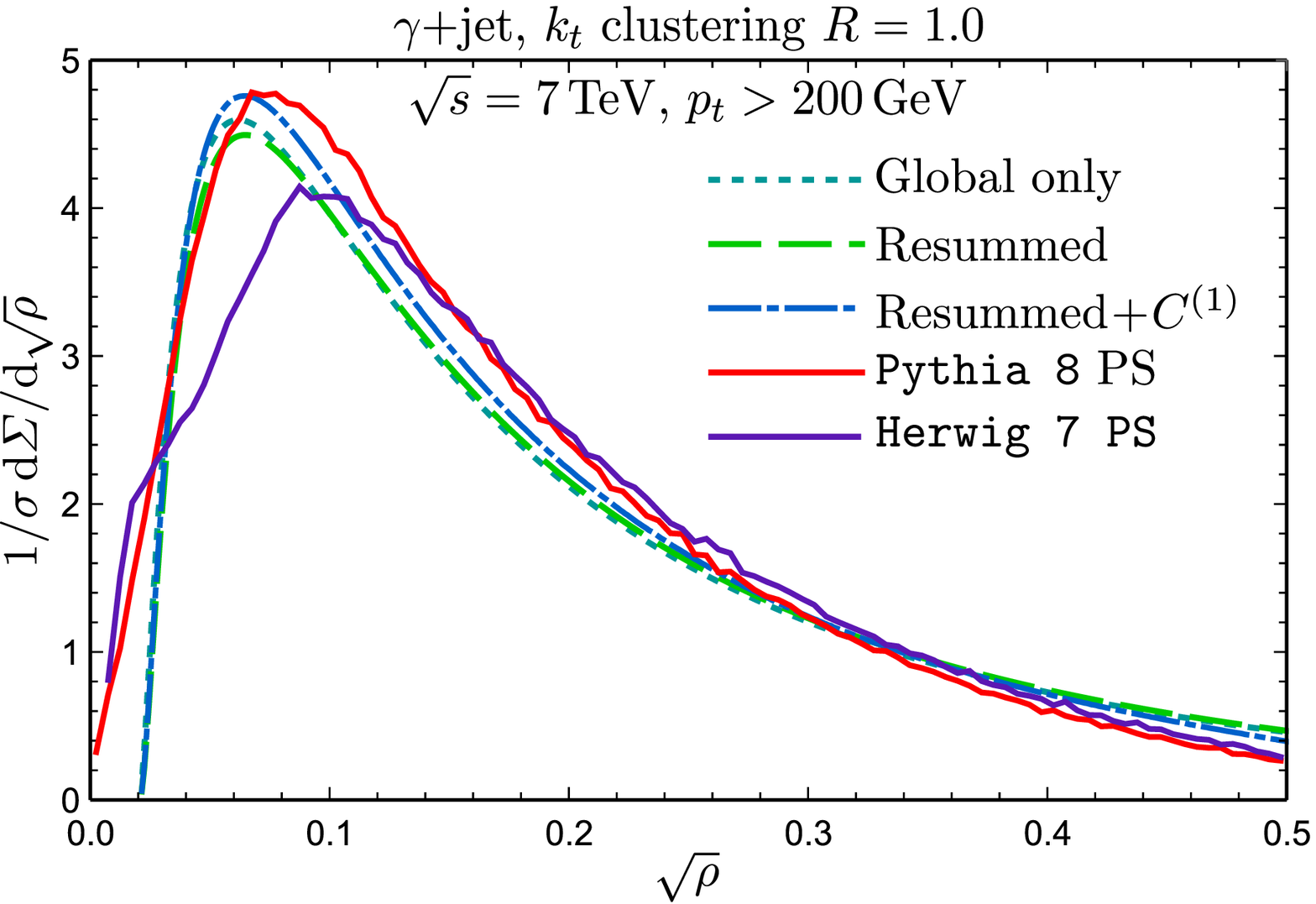}}
\caption{\label{fig:pythiaWgamma}Resummed differential jet mass distribution with $k_t$ clustering compared to \texttt{Pythia 8} and \texttt{Herwig 7} PS for $R=1.0$ in the processes $W$ + jet and $\gamma$ + jet at the LHC.}
\end{figure}

Finally, in figure \ref{fig:pythiaWgamma}, we plot the resummed differential jet mass distribution in the processes $W$ + jet and $\gamma$ + jet at the LHC with $k_t$ clustering and $R=1.0$.

Our results are in general in good agreement with \texttt{Pythia 8} results particularly near the peak of the differential distribution. The discrepancy between the results of \texttt{Pythia 8} and \texttt{Herwig 7} may be lifted when non-perturbative effects are included, as we do in the next section.

\subsection{Matching to fixed order}

\begin{figure}[ht]
\centering
\resizebox{0.48\textwidth}{!}{\includegraphics{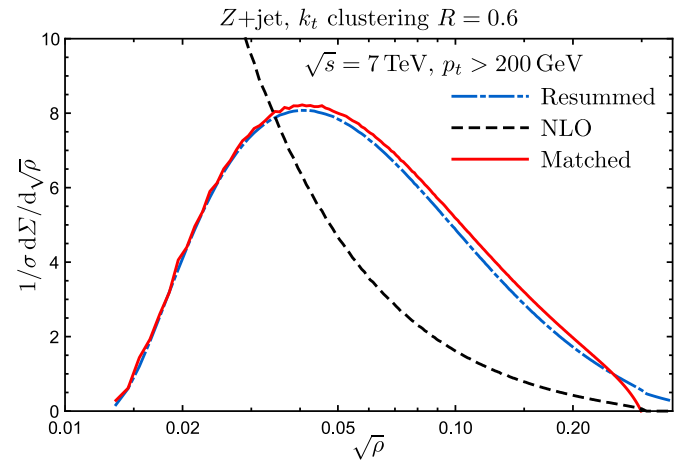}}
\caption{\label{fig:match}Matched differential jet mass distribution with $k_t$ clustering compared to \texttt{MCFM} NLO fixed-order result.}
\end{figure}

Before we end this section we discuss the matching of the resummed result with the NLO fixed-order distribution. In fact, including the constant term $C^{(1)}_{\delta}(\rho)$ in eq. \eqref{eq:dSigma}, the expansion of the resummed distribution now agrees with the fixed-order result over the entire range of $\rho$, except for the small correction due to the missing channels at the Born level (specifically the channel with incoming $gg$). Additionally, as was shown in ref. \cite{Dasgupta:2012hg}, the NLO distribution has a kinematical end point of $\rho_{\mathrm{max}}=\tan^2(R/2)$, which the resummed distribution does not have. In order to match the resummed distribution to the NLO result, specifically at the end point, we introduce the following change of the large logarithm \cite{Banfi:2010pa}
\begin{equation}
L=\ln \frac{R^2}{\rho} \to L'=\ln\left[\frac{R^2}{\rho}-\frac{R^2}{\rho_{\mathrm{max}}}+1\right],
\end{equation}
such that the large logarithm $L'$ vanishes when $\rho \to \rho_{\mathrm{max}}$, and $L'\to L$ when $\rho \to 0$. We then use the simple matching formula
\begin{equation}
\Sigma(\rho) = \Sigma_{\mathrm{NLL}}(\rho)+\Sigma_{\mathrm{NLO}}(\rho)-\Sigma_{\mathrm{NLL},\alpha_s}(\rho)\,,
\end{equation}
where now both $\Sigma_{\mathrm{NLL}}$ and $\Sigma_{\mathrm{NLL},\alpha_s}$ include the $C^{(1)}$ term. The subtracted term $\Sigma_{\mathrm{NLL},\alpha_s}$ cancels both the large logarithms and the $C^{(1)}$ terms in $\Sigma_{\mathrm{NLO}}(\rho)$, leaving only corrections due to the channels missing at the Born level. We show in figure \ref{fig:match} a plot of the matched differential jet mass distribution compared to the fixed-order result from \texttt{MCFM} for the $Z$ + jet process at the LHC. In this plot the resummed curve is plotted with the standard definition of the large logarithm ($L=\ln (R^2/\rho)$), and thus does not posses the end-point character, while the matched curve does have an end point exactly as in the \texttt{MCFM} curve. We note from this figure that the matched curve coincides with the resummed curve at small $\rho$, indicating a perfect cancellation of the large logarithms between the expanded result $\Sigma_{\mathrm{NLL},\alpha_s}$ and the fixed-order \texttt{MCFM} result $\Sigma_{\mathrm{NLO}}$.

\section{Comparison to CMS data}

In order to compare our results with the experimental data we first need to account for non-perturbative effects from hadronisation corrections and the underlying event. One commonly used numerical approach to extract these corrections is to compute the ratio of the results obtained from Monte Carlo event generators with non-perturbative effects switched on and off. In this paper we include these corrections analytically by considering the mean value of the change in the jet mass $\delta m_j^2$ due to these non-perturbative effects. This change was computed in ref. \cite{Dasgupta:2007wa} to be
\begin{equation}\label{eq:deltamj}
\langle\delta m_j^2\rangle=\sum_{(i\ell)\in\Delta_\delta}\mathcal{C}_{i\ell}\,\mathcal{A}(\mu_I)\,p_t\,\mathcal{N}_{i\ell}(R)\,,
\end{equation}
with
\begin{subequations}
\begin{align}
\mathcal{N}_{ab}(R)&=\frac{1}{4}\,R^4+\frac{1}{4608}\,R^8+\cdots\,,\\
\mathcal{N}_{aj}(R)&=\mathcal{N}_{bj}(R)=R+\frac{3}{16}\,R^3+\frac{125}{9216}\,R^5+\cdots\,,
\end{align}
\end{subequations}
and
\begin{align}
\mathcal{A}(\mu_I)&=\frac{\mu_I}{\pi}\bigg[\alpha_0(\mu_I)-\alpha_s(p_t)\notag\\
&-\frac{\beta_0}{2\pi}\left(\ln\frac{p_t}{\mu_I}+\frac{K}{\beta_0}+1\right)\alpha_s^2(p_t)\bigg].
\end{align}
Here $\mu_I$ is an arbitrary matching scale (chosen to be of order of a few GeV) and $\alpha_0$ is the averaged coupling over the non-perturbative low-$k_t$ region, $\alpha_0=1/\mu_I\int_0^{\mu_I}\alpha_s(k_t)\,\d k_t $. The $\mathcal{A}(\mu_I)$ is rescaled by the so-called Milan factor ($M=1.49$ for anti-$k_t$ clustering and $M=1.01$ for $k_t$ clustering \cite{Dasgupta:2009tm}) to account for gluon decay. The constant $K$ is defined in the appendix.

Non-perturbative effects are dominated by the contributions of the dipoles involving the outgoing jet, which scale like $\mathcal{O}(R)$, and which account for hadronisation corrections, while the smaller $\mathcal{O}(R^4)$ contributions from the incoming legs account for the underlying event. Since the mean value of $\delta m_j^2$ depends both on the Born channel and kinematics, then we perform the shift on the mass of the jet on an event-by-event basis, that is we make the change $m_j^2 \to m_j^2 - \delta m_j^2$ in the resummed form factor and then perform the convolution. Furthermore, we shift the terms $C^{(1)}(\rho)$ accordingly.

\begin{figure}[ht]
\centering
\resizebox{0.45\textwidth}{!}{\includegraphics{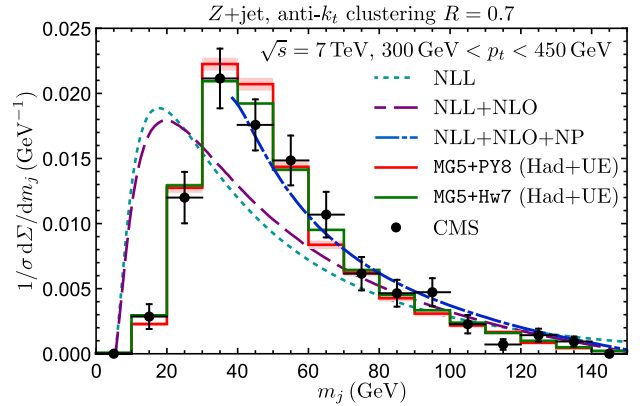}}
\caption{\label{fig:CompCMS}Resummed differential $m_j$ distribution with anti-$k_t$ clustering and $R=0.7$ in $Z(\to\ell^+\ell^-)+$ jet events, with $\ell=e,\mu$, compared to experimental data from CMS \cite{Chatrchyan:2013rla} and \texttt{MadGraph+Pythia 8} and \texttt{Herwig 7} results. The experimental data are taken from ref. \cite{1224539/t32}.}
\end{figure}

We compare, in figure \ref{fig:CompCMS}, the NLL+NLO resummed result (with the $C^{(1)}$ term), including non-perturbative corrections, with experimental data from the CMS collaboration \cite{Chatrchyan:2013rla,1224539/t32} (obtained with integrated luminosity $\mathcal{L}=5\,\mathrm{fb}^{-5}$), in the $Z$ + jet process at the LHC with anti-$k_t$ clustering and $R=0.7$. We also include in this figure the Monte Carlo results obtained from interfacing \texttt{MadGraph} with \texttt{Pythia 8} \cite{Alwall:2008qv,Conte:2012fm} and \texttt{Herwig 7} including hadronisation corrections and the underlying event. The plots in this figure are for the un-normalised jet mass variable $m_j$ rather than the normalised one $\sqrt{\rho}=m_j/p_t$. In this figure we have $300\,\mathrm{GeV}<p_t<450\,\mathrm{GeV}$. CTEQ6L parton distribution functions \cite{Pumplin:2002vw} have been used both in the convolution and \texttt{MadGraph/Pythia 8/Herwig 7/MCFM} results. For best fit we choose $\mu_I=3.5\,\mathrm{GeV}$. This plot shows a good agreement between the data and the resummed prediction over the entire range of the jet mass, as well as with the Monte Carlo simulation.

We note that the NLL+NLO+NP curve is cut off at around 40 GeV, which is just a manifestation of the shift of the resummed distribution to the right, as explained above. The value of the NLL+NLO+NP distribution at say $m_j=40$ GeV is related to the value of the resummed distribution at $\sqrt{m_j^2-\langle\delta m_j^2\rangle}$, with $\sqrt{\langle\delta m_j^2\rangle}$ (see eq. \eqref{eq:deltamj}) varying from 20 GeV to around 40 GeV depending on the $p_t$ of the jet and partonic channel. Hence, we have no result for the non-perturbative distribution below this value ($\sim40$ GeV) of the jet mass. Additionally, due to the Landau-pole singularity at small values of the jet mass, the distribution is unreliable in the region to the left of the Sudakov peak.

\section{Conclusions}

In this paper we have presented state-of-the-art detailed fixed-order calculations as well as all-orders estimates of distributions of important observables at the LHC. Specifically we have considered a typical jet-shape observable that has been studied quite substantially in the literature, namely the invariant jet mass. It is a member of a large class of observables known as non-global observables, that have so far proven to be quite delicate to treat. The subtleties in the analysis of such observables stem from the fact that they are defined for a restricted phase-space region. This is unlike global observables which are defined over the whole phase space. The former non-global observables receive contributions that are totally absent for their global counterparts. These contributions appear at each higher order in perturbation theory and have so far shown no pattern of iteration.

We have extended the work of ref. \cite{Dasgupta:2012hg} from various angles: (a) we have implemented two clustering algorithms, $k_t$ and C-A, instead of just the anti-$k_t$ considered in the latter reference. Generally, computations in $k_t$ and C-A algorithms are much more difficult to handle than in anti-$k_t$ case; (b) we have computed CLs, which are completely absent for anti-$k_t$ and thus not treated in \cite{Dasgupta:2012hg}; (c) we have investigated the jet mass distribution in various different processes, namely $W/Z/\gamma/H$ + one jet, while only $Z$ + jet was considered in the said reference; and (d) we have provided analytical expressions for our results in the form of power-series expansions in the jet radius $R$. As the experimental data \cite{Chatrchyan:2013rla} for the jet mass distribution in $Z$ + jet events at the LHC with anti-$k_t$ clustering were available only after the publication of ref. \cite{Dasgupta:2012hg}, we made the comparison of the resummed result with these experimental data herein.

We have confirmed previous results that were arrived at in studies of $e^+e^-$ annihilation processes. These include, for instance, the observation that NGLs are decreased by the application of jet clusterings other than anti-$k_t$. In other words, NGLs are more significant when anti-$k_t$ is used. This may hint at the advantage of using other jet clustering algorithms in order to bypass the difficulties posed by NGLs. Additionally, we showed that in the limit of very small jet-radius parameter the NGLs and CLs at hadron colliders coincide with those at $e^+e^-$ colliders. Moreover, we have been able to identify new features that are not present in the simple $e^+e^-$ annihilation case such as the significance of initial-state radiation and its impact on the jet mass distribution. The jet mass provides a tool to discriminate gluon and quark-initiated jets as their corresponding jet mass distributions were shown to be quite different.

It is worth, as a continuation to this project, investigating other crucial hadronic processes at the LHC such as di-jet production. The latter represents an important background for numerous potential new physics signals. Another issue that is also worth tackling is performing calculations beyond two-gluon emission. This will provide a deeper insight into the nature of QCD hadronic processes that have not been fully understood so far.

\section*{Acknowledgements}

We would like to thank Amine Ahriche and Hilal Hamdellou for assistance with \texttt{MadGraph/MadAnalysis}, and John M. Campbell for assistance with \texttt{MCFM}.

This work is supported by:
\begin{itemize}
\item Deanship of Research at the Islamic University of Madinah (research project No. 40/107)
\item PRFU: B00L02UN050120190001 (Algeria)
\end{itemize}

\appendix

\section{Born cross-section}
\label{sec:Born}

In this section we present the structure of the differential partonic Born cross-section that is needed in the convolution with the resummed result to obtain the jet mass distribution (eq. \eqref{eq:dSigma}).

The differential partonic Born cross-section (for some fixed transverse momentum $p_t$ and rapidity $y$ of the outgoing hard parton) for channel $\delta$ of the process of production of a Higgs ($H$) or a vector boson ($V$) in association with a jet in hadron collisions may be written as follows
\begin{equation}
\frac{\d\sigma_{0,\delta}}{\d\B_\delta}=\frac{1}{16\pi\,\hat{s}}\left|\mathcal{M}_{0,\delta}\right|^2\delta\left[(p_a+p_b-p_j)^2-M^2\right],
\end{equation}
where $|\mathcal{M}_{0,\delta}|^2$ is the corresponding partonic Born squared amplitude, summed and averaged over spins and colours, and $M$ is the mass of the boson. For the vector-boson processes, the differential partonic Born cross-section reduces to \cite{Ellis:1981hk,Arnold:1988dp,Gonsalves:1989ar}
\begin{equation}
\begin{split}
&q_i\bar{q}_j\to gV:\mathcal{K}_V\,\frac{2\pi\as\CF}{\Nc}\,\frac{1}{\hat{s}}\,T_0(\hat{s},\hat{u},\hat{t})\,\delta\left(\hat{s}+\hat{t}+\hat{u}-M^2\right),\\
&q_ig\to q_jV:-\mathcal{K}_V\,\frac{\pi\alpha_s}{\Nc}\,\frac{1}{\hat{s}}\,T_0(\hat{t},\hat{u},\hat{s})\,\delta\left(\hat{s}+\hat{t}+\hat{u}-M^2\right),
\end{split}
\end{equation}
where the couplings are
\begin{subequations}
\begin{align}
\mathcal{K}_Z&=\delta_{ij}\,\frac{\sqrt{2}\,\mathrm{G_F}\,M_Z^2}{4\pi}\left(g_{iV}^2+g_{iA}^2\right),\\
\mathcal{K}_W&=\left|V_{ij}\right|^2\frac{\sqrt{2}\,\mathrm{G_F}\,M_W^2}{4\pi}\,,\\
\mathcal{K}_\gamma&=\delta_{ij}\,e_i^2\,\alpha\,,
\end{align}
\end{subequations}
with $\mathrm{G_F}$ the Fermi coupling, and $g_{iV}$ and $g_{iA}$ the vector and axial-vector couplings given by
\begin{subequations}
\begin{align}
g_{iV}&=T^i_3-2\,e_i\sin^2\theta_w\,,\\
g_{iA}&=T^i_3\,,
\end{align}
\end{subequations}
where $T^i_3$ is the weak isospin of the quark $q_i$ ($+\frac{1}{2}$ for up-type and $-\frac{1}{2}$ for down-type quarks), $e_i$ is the fraction of electric charge carried by quark $q_i$, and $\theta_w$ is the Weinberg angle. Moreover, $V_{ij}$ are CKM matrix elements corresponding to flavours $i$ and $j$, with $i$ and $j$ being of different magnitude of electric charge (we exclude the top quark). In the above, $\alpha$ is the electromagnetic coupling. The kinematical factor $T_0$ describing the matrix-element squared of the underlying Born scattering is given by
\begin{equation}
T_0(\hat{s},\hat{u},\hat{t})=\frac{\hat{t}^2+\hat{u}^2+2\,\hat{s}\left(\hat{s}+\hat{t}+\hat{u}\right)}{\hat{t}\,\hat{u}}\,,
\end{equation}
with $\hat{s}$, $\hat{t}$ and $\hat{u}$ the partonic Mandelstam variables
\begin{subequations}
\begin{align}
\hat{s}&=\left(p_a+p_b\right)^2=x_a\,x_b\,s\,,\\
\hat{t}&=\left(p_a-p_j\right)^2=-x_a\,\sqrt{s}\,p_t\,e^{-y}\,,\\
\hat{u}&=\left(p_b-p_j\right)^2=-x_b\,\sqrt{s}\,p_t\,e^{y}\,.
\end{align}
\end{subequations}
For the process $gg\to H g$ (i.e. channel $(\delta_3)$) we have \cite{Ravindran:2002dc}
\begin{align}
\frac{\d\sigma_{0,\delta_3}}{\d\B_{\delta_3}}&=\frac{\mathrm{G_F}}{\sqrt{2}}\,\tau^2\,F^2\,\frac{\pi}{\CF}\left(\frac{\as}{4\pi}\right)^3\frac{1}{\hat{s}}\,\times\notag\\
&\times\frac{\hat{s}^4+\hat{t}^4+\hat{u}^4+m^8}{\hat{s}\,\hat{t}\,\hat{u}}\,\delta\left(\hat{s}+\hat{t}+\hat{u}-M_H^2\right),
\end{align}
where $\tau=7.65$ and $F=0.09$.

The total Born cross-section $\sigma_0$ is simply the integral of $\d\sigma_{0,\delta}/\d\B_{\delta}$ (including the kinematical-cuts function $\Xi_\B$) over $\B_{\delta}$, summed over possible $\delta$, where we have
\begin{equation}
\d\B_\delta=\d x_a\,\d x_b\,f_a(x_a,\mu_\mathrm{F}^2)\,f_b(x_b,\mu_\mathrm{F}^2)\,\d p_t^2\,\d y\,,
\end{equation}
where $f_i$ denotes the parton density function for the corresponding incoming parton $(i)$ evaluated at a factorisation scale $\mu_\mathrm{F}$.

Substituting the Mandelstam variables into the delta function in the integrand we obtain
\begin{equation}
\delta\left[x_a\,x_b\,s-x_a\,\sqrt{s}\,p_t\,e^{-y}-x_b\,\sqrt{s}\,p_t\,e^{y}-M^2\right].
\end{equation}
This delta function can be used to perform the integration over one of the $x$'s, say $x_b$, and thus we set
\begin{equation}
x_b=\frac{x_a\,e^{-y}\,p_t/\sqrt{s}+M^2/s}{x_a-e^{y}\,p_t/\sqrt{s}}\,,
\end{equation}
and multiply the integrand by
\begin{equation}
\Theta(1-x_b)\,\Theta(x_b)/\left(x_a\,s-\sqrt{s}\,p_t\,e^{y}\right).
\end{equation}
Since $x_b>0$ then $x_a>e^y\,p_t/\sqrt{s}$ and $y<\ln(\sqrt{s}/p_t)$. Additionally, since $x_b<1$ then $y>-\ln(\sqrt{s}/p_t)$ and
\begin{equation}\label{eq:ineq}
x_a>\frac{e^y\,p_t/\sqrt{s}+M^2/s}{1-e^{-y}\,p_t/\sqrt{s}}\,.
\end{equation}
The latter inequality overrules $x_a>e^{y}\,p_t/\sqrt{s}$, and furthermore, since $x_a<1$ we deduce that
\begin{equation}
\cosh y<\left(1-\frac{M^2}{s}\right)\frac{1}{2}\,\frac{\sqrt{s}}{p_t}\,,
\end{equation}
which also overrules the condition $|y|<\ln \sqrt{s}/p_t$.

We perform the integration over $p_t$, $y$ and $x_a$, either in the Born cross-section or in the jet mass distribution, numerically via Monte Carlo method. 

\section{Resummed Global form factor} \label{app:GlobalFormFactor}

The Sudakov global form factor that resums global logarithms is given in eq. \eqref{eq:ResummedFormFactor_global}. The radiator $\mathcal{R}$ is composed of contributions from double-logarithmic soft-collinear and single-logarithmic hard-collinear emissions from the outgoing leg $(j)$, and from single-logarithmic soft wide-angle emissions from all legs. The leading-order soft wide-angle contribution that we calculated in section \ref{sec:fo} simply exponentiates to all orders. Additionally, the non-global and clustering corrections appear as a factorised part that multiplies the resummed global form factor. The remaining soft-collinear and hard-collinear contributions may be obtained using the general formalism of resummation introduced in ref. \cite{Banfi:2004yd} as we show below.

First we write the definition of the observable $\varrho$ in terms of the transverse momentum $k_t^{(\ell)}$, rapidity $\eta^{(\ell)}$, and azimuth $\phi^{(\ell)}$ of a single soft-collinear emission with respect to the direction of the hard leg $(\ell)$. Emissions that are collinear to the incoming legs $(a)$ and $(b)$ do not end up inside the jet, so they do not contribute to its mass, hence $\varrho^{(a)}=\varrho^{(b)}=0$. For emissions that are collinear to the outgoing leg $(j)$, we introduce a coordinate rotation that takes the momentum of leg $(j)$ to the $z$ axis, where the momenta of the jet and the soft emission become
\begin{subequations}
\begin{align}
p_j^{(j)}&=p_t\cosh y\left(1,0,0,1\right),\\
k^{(j)}&=k_t^{(j)}\left(\cosh\eta^{(j)},\cos\phi^{(j)},\sin\phi^{(j)},\sinh\eta^{(j)}\right).
\end{align}
\end{subequations}
The jet mass observable (being invariant under rotations) is then given by
\begin{equation}\label{eq:nijm}
\varrho^{(j)}=2\,\frac{k^{(j)}\cdot p_j^{(j)}}{p_t^2}=2\,\frac{k_t^{(j)}}{p_t}\,e^{-\eta^{(j)}}\,\cosh y\,.
\end{equation}
Comparing the definition of the normalised invariant jet mass \eqref{eq:nijm} to the general parametrisation of observables from ref. \cite{Banfi:2004yd}
\begin{equation}\label{eq:coef}
V=d_\ell\left(\frac{k_t^{(\ell)}}{Q}\right)^{a_\ell}e^{-b_{\ell}\,\eta^{(\ell)}}\,g_{\ell}(\phi^{(\ell)})\,,
\end{equation}
where $Q=p_t$ is the hard scale, we see that $a_j = b_j = g_j=1$ and $d_j = 2\cosh y$.

Employing the master formula for resummation (eq. (3.6) from ref. \cite{Banfi:2004yd}) we obtain the expression of the radiator $\cR_\delta(\rho)$, for a given Born channel $\delta$, in the $\overline{\mathrm{MS}}$ renormalisation scheme
\begin{align}\label{eq:master}
\mathcal{R}_\delta(\rho)&=C_j\left[L\,g_1(\alpha_sL)+g_2(\alpha_sL)+g_{2,\mathrm{coll}}(\alpha_sL)\right]+\notag\\
&+g_{2,\mathrm{wide}}(\alpha_sL)\left[\mathcal{C}_{ab}\,\frac{R^2}{2}+\left(\mathcal{C}_{aj}+\mathcal{C}_{bj}\right)h(R)\right],
\end{align}
with $h(R)$ being given in eq. \eqref{eq:fR}. Here $C_j$ is the colour factor associated with leg $j$, $C_j=\CF$ for outgoing (anti-) quark jet and $C_j=\CA$ for outgoing gluon jet, and $\mathcal{C}_{i\ell}$ is the colour factor for dipole $(i\ell)$ introduced in the main text. We have
\begin{subequations}
\begin{align}
&g_1=\frac{1}{2\pi\beta_0\lambda}\left[(1-2\lambda)\ln(1-2\lambda)-2(1-\lambda)\ln(1-\lambda)\right],\\
&g_2=\frac{\mathrm{K}}{4\pi^2\beta_0^2}\left[2\ln(1-\lambda)-\ln(1-2\lambda)\right]+\frac{\beta_1}{2\pi\beta_0^3}\times\notag\\
&\left[\frac{1}{2}\ln^2(1-2\lambda)-\ln^2(1-\lambda)+\ln(1-2\lambda)-2\ln(1-\lambda)\right],\\
&g_{2,\mathrm{coll}}=-B_j\,\frac{1}{\pi\beta_0}\,\ln(1-\lambda)\,,\\
&g_{2,\mathrm{wide}}=-\frac{1}{2\pi\beta_0}\ln(1-2\lambda)\,,
\end{align}
\end{subequations}
with $\lambda=\as(R\,p_t)\,\beta_0\,\ln(R^2/\rho)$. The factor $B_j$ accounts for corrections due to hard-collinear emissions to the outgoing jet $j$ and is given by
\begin{equation}
\begin{aligned}
B_q &= -\frac{3}{4}                                           &&\text{for quark jets}\,, \\
B_g &= -\frac{11\,\CA-4\,\mathrm{T_R}\,\mathrm{n_f}}{12\,\CA}=-\frac{\pi\beta_0}{\CA} \quad&&\text{for gluons jets}\,,
\end{aligned}
\end{equation}
with $\mathrm{T_R}$ the Dynkin index (normalisation constant) for the $\mathrm{SU}(\Nc)$ generators, $\mathrm{T_R}=1/2$, and $\mathrm{n_f}=5$ the number of active quark flavours. Additionally we have
\begin{equation}
\begin{split}
\mathrm{K}&= \CA\left(\frac{67}{18}-\frac{\pi^2}{6}\right)-\frac{5}{9}\,\mathrm{n_f}\,,\\
\beta_0&=\frac{11\,\mathrm{C_A}-2\,\mathrm{n_f}}{12\,\pi}\,,\\
\beta_1&=\frac{17\,\mathrm{C_A^2}-5\,\mathrm{C_A}\,\mathrm{n_f}-3\,\mathrm{C_F}\,\mathrm{n_f}}{24\pi^2}\,.
\end{split}
\end{equation}
In the master formula we excluded the single-logarithmic soft wide-angle term referred to in ref. \cite{Banfi:2004yd} as $\ln S(T)$, and we calculated it manually in section \ref{sec:fo}. It appears as the last term in the radiator \eqref{eq:master}.

The derivative of the radiator $\mathcal{R}$ with respect to $L$, relevant in the expression \eqref{eq:ResummedFormFactor_global}, is given by
\begin{equation}
\mathcal{R}'=\frac{\partial\mathcal{R}}{\partial L}=\frac{C_j}{\pi\beta_0}\left[\ln(1-\lambda)-\ln(1-2\lambda)\right].
\end{equation}

\section{Fixed-order expansion}

For the sake of matching we need the fixed-order expansion of the resummed form factor \eqref{eq:all}. We can cast the latter in the form
\begin{align}
\frac{\d\Sigma_\delta(\rho)}{\d\B_{\delta}}&=\frac{\d\sigma_{0,\delta}}{\d\B_{\delta}}\,C_\delta(\rho)\,\exp\left[\sum_{n=1}^\infty\sum_{m=1}^{n+1}G_{nm}\,\bar{\alpha}_s^n\,L^m \right]\notag\\
&=\frac{\d\sigma_{0,\delta}}{\d\B_{\delta}}\,\sum_{n=0}^\infty\sum_{m=0}^{2n} H_{nm} \, \bar{\alpha}_s^n\,L^m,
\end{align}
where the expansion coefficients in the exponent, $G_{nm}$, up to $\cO(\asb^2)$, are
\begin{align}
G_{12}&=-\frac{C_j}{2}\,,\notag \\
G_{11}&=-B_j\,C_j-\cC_{ab}\,\frac{R^2}{2}-\left(\cC_{aj}+\cC_{bj}\right)h(R)\,,\notag\\
G_{23}&=-\frac{\pi\beta_0}{2}\,C_j\,,\notag \\
G_{22}&=-\frac{C_j}{4}\left(K+2\pi\beta_0\,B_j\right)-\pi\beta_0\left(\cC_{aj}+\cC_{bj}\right)h(R)+\notag\\
&-\pi\beta_0\,\cC_{ab}\,\frac{R^2}{2}+\frac{\cF_2^{\delta}}{2}-\frac{\cS_2^{\delta}}{2}-\frac{\zeta_2}{2}\,.
\end{align}
and the expansion coefficients in the series, $H_{nm}$, are
\begin{align}
&H_{12}=G_{12}\,,\qquad H_{11}=G_{11}\,,\qquad H_{10}=C^{(1)}_{\delta}\,,\notag\\
&H_{24}=\frac{G_{12}^2}{2}, \qquad H_{23} = G_{23} + G_{12}\, G_{11},\notag\\
&H_{22}=G_{22}+\frac{G_{11}^2}{2}+\left(C^{(1)}_{\delta}\right)G_{12}.
\end{align}


\end{document}